\documentclass[reqno]{amsart}
\usepackage{amsmath,amssymb,amsthm}
\usepackage{graphicx,a4wide}

\theoremstyle{plain}

\theoremstyle{remark}

\newcommand{\prn}[1]{\left(#1\right)}
\newcommand{\brk}[1]{\left[#1\right]}
\newcommand{\pair}[2]{\left<#1,#2\right>_{SS'}}
\newcommand{\R}{\mathbb{R}}
\newcommand{\E}[1]{\mathbb{E}\left[#1\right]}
\renewcommand{\vec}[1]{\mathbf{#1}}
\newcommand{\mat}[1]{\mathbf{#1}}
\newcommand{\ud}[1]{\, \mathrm{d}#1}

\begin{document}
\parindent0ex
\parskip2ex

\title[Optimal prediction for moment models]
{Optimal prediction for moment models: Crescendo diffusion and reordered equations}

\author{Benjamin Seibold}
\address{Massachusetts Institute of Technology \\
77 Massachusetts Avenue \\ Cambridge, MA 02139}
\email{seibold@math.mit.edu}
\urladdr{http://www-math.mit.edu/\~{ }seibold}

\author{Martin Frank}
\address{TU Kaiserslautern,
Erwin Schr{\"o}dinger Strasse \\ D-67663 Kaiserslautern \\ Germany}
\email{frank@mathematik.uni-kl.de}
\urladdr{http://www.mathematik.uni-kl.de/\~{ }frank}


\thanks{The authors thank Martin Grothaus for helpful suggestions on measures in
function spaces. The support by the German Research Foundation and
the National Science Foundation is acknowledged.
M.~Frank was supported by DFG grant KL 1105/14/2.
B.~Seibold was partially supported by NSF grant DMS--0813648.
}

\begin{abstract}
A direct numerical solution of the radiative transfer equation or any kinetic equation
is typically expensive, since the radiative intensity depends on time, space and direction.
An expansion in the direction variables yields an equivalent system of infinitely many
moments. A fundamental problem is how to truncate the system. Various closures have been
presented in the literature. We want to generally study moment closure within the
framework of optimal prediction, a strategy to approximate the mean solution of a large
system by a smaller system, for radiation moment systems. We apply this strategy to
radiative transfer and show that several closures can be re-derived within this framework,
such as $P_N$, diffusion, and diffusion correction closures. In addition, the formalism gives
rise to new parabolic systems, the reordered $P_N$ equations, that are similar to the
simplified $P_N$ equations. Furthermore, we propose a modification to existing closures.
Although simple and with no extra cost, this newly derived crescendo diffusion yields
better approximations in numerical tests.
\end{abstract}

\maketitle

\section{Introduction}
In many fields, macroscopic equations can be derived from mesoscopic kinetic equations.
For instance, in the Navier-Stokes and Euler equations, the macroscopic fluid variables,
e.g.~density and momentum, are moments of the phase space distribution of the Boltzmann
equation. Similarly, in the equations of radiative transfer \cite{Modest1993}, the
direction dependent kinetic equations can be transformed into a coupled system for the
infinitely many moments. This can be interpreted as considering an (infinite) expansion
in the kinetic variable and then taking only finitely many members of this expansion.
This usually means that several ``coordinates'' of the field are neglected.

A common feature of existing closure strategies is that they are based on truncating
and approximating the moment equations, and observing to which extent the solution of
the approximate system is close to the true solution. The approximations are supported
by physical arguments, such as higher moments being small or adjusting instantaneously.
In Sect.~\ref{sec:moment_system_radiation}, we derive a moment model for radiative
transfer, and in Sect.~\ref{sec:moment_closure}, we outline some classical linear closure
strategies.

A more potent strategy would be to have an identity for the evolution of the first
$N$ moments, and then deriving closures by approximating this identity. Moment models
have first been studied from a higher perspective by Levermore \cite{Levermore2005}.
Recent systematic studies of moment systems include also Struchtrup's order of magnitude
method \cite{Struchtrup2004} which leads to the R13 equations of
gas dynamics \cite{StruchtrupTorrilhon2003,StruchtrupTorrilhon2008}. In this paper,
we take a similar approach. Extending
results of \cite{FrankSeibold2008}, we show that the method of optimal prediction
\cite{ChorinKastKupferman1998_1,ChorinHaldKupferman2000,ChorinHaldKupferman2002,
Chorin2003,ChorinHald2006}
can be applied to the equations of radiative transfer and yields closed systems of
finitely many moments. Optimal prediction, outlined in Sect.~\ref{sec:optimal_predition},
approximates the mean solution of a large system by a smaller
system, by averaging the equations with respect to an underlying probability measure.
It can be understood as removing undesired modes, but in an averaged fashion, instead
of merely neglecting them.

Optimal prediction has been formulated for Hamiltonian partial differential
equations \cite{ChorinKastKupferman1998_1,ChorinKastKupferman1998_2}, however, without
exploiting the full formalism. It has been applied to partial differential
equations \cite{ChorinKastKupferman1999,BellChorinCrutchfield2000}, however, only after
reducing them to a system of ordinary differential equations using a Fourier expansion
or a semi-discretization step. In addition, most considered examples are Hamiltonian,
for which a canonical measure exists.
In contrast, here we encounter partial differential equations (in space) after a Fourier
expansion (in the angular variable). Hence, the methodology is generalized to
semigroups. Furthermore, the radiation system is linear. Using Gaussian measures, the
formalism is linear, and it yields an identity in the presence of a memory kernel.
Here we restrict ourselves to Gaussian measures with vanishing covariance.
We present linear optimal prediction in function spaces, and show various possible
approximations of the memory term.

In Sect.~\ref{sec:apply_op_radiation}, we apply linear optimal prediction to the
radiation moment system, and derive existing and propose new closure relations.
The new formalism allows a better understanding of the errors done due to the
truncation of the infinite system. The performance of the newly derived closures is
investigated numerically in Sect.~\ref{sec:numerical_results}.

\section{Moment Models for Radiative Transfer}
\label{sec:moment_system_radiation}
The equations of radiative transfer \cite{Modest1993} are
\begin{equation}
\frac{1}{c}\partial_t I(x,\Omega,t)+\Omega\nabla I(x,\Omega,t)
+(\sigma(x)+\kappa(x)) I(x,\Omega,t)
= \frac{\sigma(x)}{4\pi}\int_{4\pi}I(x,\Omega',t) d\Omega'+q(x,t)\;.
\label{eq:radiative_equation_full}
\end{equation}
In these equations, the radiative intensity $I(x,\Omega,t)$ can be viewed as the
the number of particles at time $t$, position $x$, traveling into
direction $\Omega$. Equation \eqref{eq:radiative_equation_full} is a mesoscopic
phase space equation, modeling absorption and emission ($\kappa$-term), scattering
($\sigma$-term), and containing a source term $q$. Due to the large number of unknowns,
a direct numerical simulation
of \eqref{eq:radiative_equation_full} is very costly. Often times only the lowest
moments of the intensity with respect to the direction $\Omega$ are of interest. Moment
models attempt to approximate \eqref{eq:radiative_equation_full} by a coupled system of
moments.

For the sake of notational simplicity, we consider a slab geometry. However, all methods
presented here can be easily generalized. Consider a plate that is finite along the
$x$-axis and infinite in the $y$ and $z$ directions. The system is assumed to be invariant
under translations in $y$ and $z$ and in rotations around the $x$-axis. In this case the
radiative intensity can only depend on the scalar variable $x$ and on the azimuthal angle
$\theta = \arccos(\mu)$ between the $x$-axis and the direction of motion. Furthermore,
we select units such that $c=1$. The system becomes
\begin{equation}
\partial_t I(x,\mu,t) + \mu \partial_x I(x,\mu,t)  +(\sigma(x)+\kappa(x)) I(x,\mu,t)
=\frac{\sigma(x)}{2}\int_{-1}^1 I(x,\mu',t)\ud{\mu'}+q(x,t)
\label{eq:radiation_equation_simple}
\end{equation}
with $t>0$, $x\in]a,b[$, and $\mu\in[-1,1]$. The system is supplied with boundary
conditions that either prescribe ingoing characteristics or are periodic boundary
conditions
\begin{equation*}
\begin{cases} I(a,\mu,t)\vert_{\mu>0} = 0 \\ I(b,\mu,t)\vert_{\mu<0} = 0 \end{cases}
\quad\text{or}\quad
\begin{cases} I(a,\mu,t) = I(b,\mu,t) \\ I_x(a,\mu,t) = I_x(b,\mu,t) \end{cases}
\end{equation*}
and initial conditions
\begin{equation*}
I(x,\mu,0) = \mathring{I}(x,\mu).
\end{equation*}
Under very general assumptions, this problem admits a unique solution
$I \in C([0,\tau];L^2([a,b]\times[-1,1]))$,
i.e.~at each time $I$ is square-integrable in both spatial and angular
variables \cite{DautrayLions1993}.

Macroscopic approximations can be derived using angular basis functions. Commonly used
are Legendre polynomials \cite{Davison1958,Chandrasekhar1960}, since they form an
orthogonal basis on $[-1,1]$. Define the moments
\begin{equation*}
I_l(x,t) = \int_{-1}^1 I(x,\mu,t)P_l(\mu)\ud{\mu}.
\end{equation*}
Multiplying (\ref{eq:radiation_equation_simple}) with $P_k$ and integrating over $\mu$
gives
\begin{equation*}
\partial_t I_k(x,t) + \partial_x \int_{-1}^1 \mu P_k(\mu) I(x,\mu,t)d\mu
+(\sigma(x)+\kappa(x)) I_k(x,t)
= (\sigma(x)I_0(x,t)+2q(x,t))\delta_{k,0}.
\end{equation*}
Using the recursion relation for the Legendre polynomials leads to
\begin{equation*}
\partial_t I_k+\partial_x \sum_{l=0}^\infty b_{kl} I_l
= \begin{cases} \kappa (2q-I_0), & k=0 \\ -(\kappa+\sigma) I_k, & k>0 \end{cases}
\end{equation*}
where $b_{kl} = \tfrac{k+1}{2k+1}\delta_{k+1,l}+\tfrac{k}{2k+1}\delta_{k-1,l}$.
This is an infinite tridiagonal system for $k=0,1,2,\dots$
\begin{equation}
\partial_t I_k+b_{k,k-1}\partial_x I_{k-1}+b_{k,k+1}\partial_x I_{k+1} = -c_k I_k+q_k\;,
\label{eq:radiative_moment_system}
\end{equation}
of first-order partial differential equations.
Using the (infinite) matrix notation
\begin{equation}
\mat{B} = \begin{pmatrix}
0 & 1 & & & \\
\frac{1}{3} & 0 & \frac{2}{3} & & \\
& \frac{2}{5} & 0 & \frac{3}{5} & \\
& & \frac{3}{7} & 0 & \ddots \\ & & & \ddots & \ddots
\end{pmatrix}
\quad\text{,}\quad
\mat{C} = \begin{pmatrix}
\kappa & & & \\
& \kappa+\sigma & & \\
& & \kappa+\sigma & \\
& &  & \ddots
\end{pmatrix}
\quad\text{and}\quad
\mat{q} = \begin{pmatrix} 2\kappa q \\ 0 \\ \vdots \\ \vdots \end{pmatrix}
\label{eq:radiative_moment_matrices}
\end{equation}
we can write \eqref{eq:radiative_moment_system} as
\begin{equation}
\label{eq:radiative_equation_matrix}
\partial_t\vec{I}+\vec{B}\cdot\partial_x\vec{I} = -\vec{C}\cdot\vec{I}+\vec{q}\;.
\end{equation}
The infinite moment system \eqref{eq:radiative_moment_system} is equivalent to the
transfer equation \eqref{eq:radiation_equation_simple}, since we represented an $L^2$
function in terms of its Fourier components. In order to admit a numerical computation,
\eqref{eq:radiative_moment_system} has to be approximated by a system of finitely many
moments $I_0,\dots,I_N$, i.e.~all modes $I_l$ with $l>N$ are not considered.
Mathematically, it is evident that such a truncation approximates the full system,
since the $I_l$ decay faster than $\frac{1}{l}$, due to
\begin{equation*}
\|I(x,\cdot,t)\|_{L^2(-1,1)}^2 = \sum_{l=0}^\infty\tfrac{2l+1}{2}I_l(x,t)^2 < \infty\;.
\end{equation*}

\section{Moment Closure}
\label{sec:moment_closure}
In order to obtain a closed system, in the equation for $I_N$, the dependence on $I_{N+1}$
has to be eliminated. A question of fundamental interest is how to close the moment system,
i.e.~by what to replace the dependence on $I_{N+1}$. In the following, we name three types
of linear closure approaches: the $P_N$ closure, higher-order diffusion (correction)
approximations, and the simplified $P_N$ closure.

\subsection{$P_N$ closure}
\label{subsec:closure_PN}
The simplest closure, the so-called \emph{$P_N$ closure} \cite{Chandrasekhar1944}
is to truncate the sequence $I_l$, i.e.~$I_l=0$ for $l>N$.
The physical argument is that if the system is close to equilibrium, then the underlying
particle distribution is uniquely determined by the lowest-order moments. This can be
justified rigorously by an asymptotic analysis of Boltzmann's equation \cite{Caflisch1980}.

\subsection{Diffusion correction closures}
\label{subsec:closure_diffusion}
The classical diffusion closure is defined for $N=1$. We assume $I_1$ to be
quasi-stationary and neglect $I_l$ for $l>1$, thus the equations read
\begin{align*}
\partial_t I_0 + \partial_x I_1 &= -\kappa I_0+q_0 \\
\tfrac{1}{3}\partial_x I_0 &= - (\kappa+\sigma) I_1\;.
\end{align*}
Solving the second equation for $I_1$ and inserting it into the first equation yields
the diffusion approximation
\begin{equation}
\partial_t I_0 - \partial_x \tfrac{1}{3(\sigma+\kappa)}\partial_x I_0 = -\kappa I_0+q_0\;.
\label{eq:closure_diffusion}
\end{equation}

A new hierarchy of $P_N$ approximations, denoted \emph{diffusion correction} or
\emph{modified diffusion} closure, has recently been proposed by
Levermore \cite{Levermore2005}. In slab geometry, it can be derived in the following
way: We assume that $I_l=0$ for $l>N+1$. Contrary to the $P_N$ closure, the $(N+1)$-st
moment is assumed to be quasi-stationary. Setting $\partial_t I_{N+1} = 0$ yields the
algebraic relation
\begin{equation*}
I_{N+1} = -\tfrac{1}{\kappa+\sigma}\tfrac{N+1}{2N+3}\partial_x I_N\;,
\end{equation*}
which, substituted into the equation for $I_N$, yields an additional diffusion term
for the last moment:
\begin{equation}
\partial_t I_N+b_{N,N-1}\partial_x I_{N-1}
-\tfrac{1}{\kappa+\sigma}\vartheta_N\partial_{xx} I_N
= \begin{cases} \kappa (2q-I_0), & N=0 \\ -(\kappa+\sigma) I_N, & N>0 \end{cases}
\label{eq:closure_diffusion_correction}
\end{equation}
where $\vartheta_N = b_{N,N+1}\frac{N+1}{2N+3} = \frac{(N+1)^2}{(2N+1)(2N+3)}$.
For $N=0$ this closure becomes the classical diffusion
closure \eqref{eq:closure_diffusion}.

\subsection{Simplified $P_N$ closure}
\label{subsec:closure_SPN}
Other higher order diffusion approximations exist, such as the so-called
\emph{simplified $P_N$} ($SP_N$) equations. For the steady case, these have been derived in
an ad hoc fashion \cite{Gelbard1960,Gelbard1961,Gelbard1962} and have subsequently been
substantiated via asymptotic analysis \cite{LarsenMorelMcGhee1996} and via a variational
approach \cite{TomasevicLarsen1996,BrantleyLarsen2000}. In the time-dependent case, the
simplified $P_N$ equations have been derived by asymptotic
analysis \cite{FrankKlarLarsenYasuda2007}.
They are a system of parabolic partial differential equations. The $SP_1$ equations are
identical to the diffusion approximation. The $SP_3$ equations are
\begin{align*}
\partial_t\phi   &= \frac{1}{3\sigma_t}\nabla_x^2
\brk{\phi+2\phi_2-\zeta}-\sigma_a\phi+q \\
\partial_t\phi_2 &= \frac{1}{3\sigma_t}\nabla_x^2
\brk{\frac{2}{15\alpha}\phi+\frac{11}{21\alpha}\phi_2}-\frac{1}{3\alpha}\sigma_t\phi_2 \\
\partial_t\zeta  &= \frac{1}{3\sigma_t}\nabla_x^2
\brk{\phi+2\phi_2+\prn{\frac{12}{5}(1-\alpha)-1}\zeta}-\sigma_a\phi+q-\sigma_t\zeta\;.
\end{align*}
In these equations, $\alpha$ is a free parameter. To obtain a well-posed parabolic system, it should be set \cite{FrankKlarLarsenYasuda2007} to $0<\alpha\lesssim 0.9$. There are two particularly sensible choices:
For $\alpha=\frac{1}{3}$, the equations reduce to the steady-state $SP_3$ equations, whereas for $\alpha=\frac{2}{3}$, they are asymptotically close to the time-dependent $P_3$ equations. In the following, we set $\alpha$ to $\frac{1}{3}$.

These equations can be approximated by a quasi-steadiness assumption. This gives the
simplified simplified $P_3$ ($SSP_3$) approximation
\begin{align*}
\partial_t\phi   &= \frac{1}{3\sigma_t}\nabla_x^2
\brk{\phi+2\phi_2}-\sigma_a\phi+q \\
\partial_t\phi_2 &= \frac{1}{3\sigma_t}\nabla_x^2
\brk{\frac{2}{15\alpha}\phi+\frac{11}{21\alpha}\phi_2}-\frac{1}{3\alpha}\sigma_t\phi_2\;.
\end{align*}

\subsection{Other types of closures}
Various nonlinear approximations exist in the literature, most prominently flux-limited
diffusion \cite{Levermore1984} and minimum entropy methods \cite{MullerRuggeri1993,
AnilePennisiSammartino1991,DubrocaFeugeas1999,TurpaultFrankDubrocaKlar2004,
FrankDubrocaKlar2006}.

\section{Optimal Prediction}
\label{sec:optimal_predition}
Optimal prediction, as introduced by Chorin, Hald, Kast, Kupferman et~al.
\cite{ChorinKastKupferman1998_1,ChorinHaldKupferman2000,ChorinHaldKupferman2002,
Chorin2003,ChorinHald2006}
seeks the mean solution of a time-dependent system, when only part of the initial data
is known. It is assumed that a measure on the phase space is available. Fundamental
interest lies in nonlinear systems, for which the mean solution decays to a
thermodynamical equilibrium. Applications include molecular dynamics \cite{Seibold2004}
and finance \cite{Okunev2005}. The formalism has been developed in detail
\cite{ChorinHaldKupferman2000,ChorinHaldKupferman2001,ChorinHald2006}
for dynamical systems
\begin{equation}
\vec{\dot x}(t) = \vec{R}(\vec{x}(t)) \ , \quad \vec{x}(0) = \vec{\mathring{x}}\;.
\label{eq:ode_system}
\end{equation}
Let the vector of unknowns be split $\vec{x} = (\hat{\vec{x}},\tilde{\vec{x}})$ into a
part of interest $\hat{\vec{x}}$, for which the initial conditions
$\mathring{\hat{\vec{x}}}$ are known, and a part to be ``averaged out'' $\tilde{\vec{x}}$,
for which the initial conditions $\mathring{\tilde{\vec{x}}}$ are not known or not of
relevance. In addition, let a probability measure $f(\vec{x})$ be given (for Hamiltonian
systems this could be the grand canonical distribution
$f(\vec{x}) = Z^{-1}\exp\prn{-\beta H(\vec{x})}$). Given the known initial conditions
$\mathring{\hat{\vec{x}}}$, the measure $f$ induces a conditioned measure
$f_{\vec{\mathring{\hat{x}}}}(\tilde{\vec{x}}) =
\tilde Z^{-1} f(\mathring{\hat{\vec{x}}},\tilde{\vec{x}})$ for the remaining unknowns.
An average of a function $u(\hat{\vec{x}},\tilde{\vec{x}})$ with respect to
$f_{\vec{\mathring{\hat{x}}}}$ is the conditional expectation
\begin{equation}
Pu = \E{u|\hat{\vec{x}}} = \frac
{\int u(\hat{\vec{x}},\tilde{\vec{x}})f(\hat{\vec{x}},\tilde{\vec{x}})\ud{\tilde{\vec{x}}}}
{\int f(\hat{\vec{x}},\tilde{\vec{x}})\ud{\tilde{\vec{x}}}}\;.
\label{eq:cond_expect_operator}
\end{equation}
It is an orthogonal projection with respect to the inner product $(u,v) = \E{uv}$,
which is defined by the expectation $\E{u} = \int\int u(\hat{\vec{x}},\tilde{\vec{x}})
f(\hat{\vec{x}},\tilde{\vec{x}})\ud{\hat{\vec{x}}}\ud{\tilde{\vec{x}}}$.
Let $\varphi(\vec{x},t)$ denote the solution of \eqref{eq:ode_system}, for the initial
conditions $\vec{x}=(\hat{\vec{x}},\tilde{\vec{x}})$. Then optimal prediction seeks for
the mean solution
\begin{equation}
P\varphi(\vec{x},t) = \E{\varphi(\hat{\vec{x}},\tilde{\vec{x}},t)|\hat{\vec{x}}}\;.
\label{eq:op_mean_solution}
\end{equation}
A possible, but computationally expensive approach to approximate
\eqref{eq:op_mean_solution} is by Monte-Carlo sampling, as presented in
\cite{ChorinHaldKupferman2002}.
Optimal prediction formulates a smaller system for $\hat{\vec{x}}$.
First order optimal prediction \cite{ChorinKastKupferman1999} constructs this system by
applying the conditional expectation \eqref{eq:cond_expect_operator}
to the original equation's \eqref{eq:ode_system} right hand side.
For Hamiltonian systems, the arising system is again Hamiltonian \cite{ChorinHald2006}.

An approximate formula for the mean solution can be derived by applying the
Mori-Zwanzig formalism \cite{Mori1965,Zwanzig1973} in a version for conditional
expectations \cite{ChorinHaldKupferman2000} to the Liouville equation for
\eqref{eq:ode_system}. It yields an integro-differential equation, that involves the
first order right hand side, plus a memory kernel. First order optimal
prediction can be interpreted as the crude approximation of dropping the memory term.
Various better approximations have been presented in
\cite{ChorinHaldKupferman2001,ChorinHaldKupferman2002,Chorin2003}.
For nonlinear systems, optimal prediction remains an approximation, even if the memory
term is computed exactly.

\subsection{Linear Optimal Prediction}
\label{subsec:linear_optimal_prediction}
Consider a linear system of evolution equations
(such as \eqref{eq:radiative_equation_matrix})
\begin{equation}
\partial_t\vec{u} = R\vec{u} \ , \quad \vec{u}(0) = \vec{\mathring{u}}\;,
\label{eq:linear_system}
\end{equation}
where $R$ is a differential operator. Let the unknowns and the operator be split
\begin{equation*}
\vec{u} = \begin{bmatrix} \hat{\vec{u}} \\ \tilde{\vec{u}} \end{bmatrix}
\quad\text{and}\quad
R = \begin{bmatrix}
\hat{\hat{R}} & \hat{\tilde{R}} \\
\tilde{\hat{R}} & \tilde{\tilde{R}}
\end{bmatrix}.
\end{equation*}

Lebesgue measures cannot be generalized to infinite dimensions.
Instead, as in \cite{HidaKuoPotthoffStreit1993,BerezanskyKondratiev1995},
we define measures on function spaces as follows. Let $S$ be a space of test functions
(e.g.\ the Schwartz space) and let $S'$ be the corresponding space of distributions, such
that $S\subset L^2\subset S'$ is a Gelfand triple \cite{BerezanskyKondratiev1995}.
Measures on vector valued function spaces are defined via a vector valued Gelfand
triple $S^n\subset (L^2)^n\subset (S')^n$.

Let $\vec{f}\in S^n$ be a vector-valued test function, $\vec{u}\in (S')^n$ a
vector-valued distribution, and $\mat{\Sigma}\in (S')^{n\times n}$ a
matrix-valued distribution. Then the pair-dot-product
\begin{equation*}
\pair{\vec{f}}{\vec{u}} = \sum_j \pair{f_j}{u_j} \in\R
\end{equation*}
is a number, and the pair-matrix-vector-product
\begin{equation*}
\pair{\vec{f}}{\mat{\Sigma}}
= \begin{pmatrix}
\sum_j \pair{f_j}{\Sigma_{j1}} \\ \vdots \\ \sum_j \pair{f_j}{\Sigma_{jn}}
\end{pmatrix}
\in\R^n
\end{equation*}
is a vector.

We define a Gaussian measure by its Fourier transform
\begin{equation}
\int \exp\prn{i\pair{\vec{f}}{\vec{u}}}\ud{\lambda(\vec{u})}
= \exp\prn{-\frac{\|\pair{\vec{f}}{\mat{\Sigma}}\|_2^2}{2}+i\pair{\vec{f}}{\vec{m}}}\;,
\label{eq:Gaussian_measure_function}
\end{equation}
where the variance $\mat{\Sigma}$ is a matrix-valued function and the expectation
value $\mat{m}$ is a vector-valued distribution.

Here we assume a measure with vanishing covariance between the different moments
(i.e. diagonal $\mat{\Sigma}$).
As derived in \cite{FrankSeibold2008}, this leads to the simple linear projections
\begin{equation*}
P\vec{x} = \mat{E}\vec{x} \quad\text{and}\quad (I-P)\vec{x} = \mat{F}\vec{x}
\end{equation*}
with
\begin{equation*}
\mat{E} = \begin{bmatrix} \mat{I} & \mat{0} \\ \mat{0} & \mat{0} \end{bmatrix}
\quad\text{and}\quad
\mat{F} = \begin{bmatrix} \mat{0} & \mat{0} \\ \mat{0} & \mat{I} \end{bmatrix}\;.
\end{equation*}
Considering the solution operators $e^{tR}$ and $e^{tR\mat{F}}$, which we assume
to be well posed, the Mori-Zwanzig formalism \cite{Mori1965,Zwanzig1973} yields the
identity
\begin{equation}
\partial_t\prn{e^{tR}}
= \mathcal{R}e^{tR}+e^{tR\mat{F}}R\mat{F}+\int_0^t K(t-s)e^{sR}\ud{s}\;,
\label{eq:solution_operator_identity}
\end{equation}
where
\begin{equation}
\mathcal{R} = R\mat{E}
\label{eq:projected_rhs}
\end{equation}
is the projected differential operator, and
\begin{equation}
K(t) = e^{tR\mat{F}}R\mat{F}R\mat{E}
\label{eq:memory_kernel}
\end{equation}
is a memory kernel for the dynamics.
The projected operator $e^{tR}\mat{E}$ represents the mean solution in the context of
optimal prediction. Its evolution is given by
\begin{equation}
\partial_t\prn{e^{tR}\mat{E}} = \mathcal{R}e^{tR}\mat{E}
+\int_0^t K(t-s)e^{sR}\mat{E}\ud{s}\;.
\label{eq:projected_solution_operator_identity}
\end{equation}
Compared to \eqref{eq:solution_operator_identity}, the middle term has canceled,
since $\mat{F}\mat{E} = \mat{0}$. The projected differential operators, the
projected right hand side \eqref{eq:projected_rhs} and
the memory kernel \eqref{eq:memory_kernel} have the block form
\begin{equation*}
\mathcal{R} = R\mat{E}
= \begin{bmatrix} \hat{\hat{R}} & 0 \\ \tilde{\hat{R}} & 0 \end{bmatrix}
\quad\text{,}\quad
R\mat{F}
= \begin{bmatrix} 0 & \hat{\tilde{R}} \\ 0 & \tilde{\tilde{R}} \end{bmatrix}
\quad\text{,}\quad
R\mat{F}R\mat{E} =
\begin{bmatrix} \hat{\tilde{R}}\tilde{\hat{R}} & 0 \\
\tilde{\tilde{R}}\tilde{\hat{R}} & 0 \end{bmatrix}
\quad\text{and}\quad
K = \begin{bmatrix} \hat{\hat{K}} & 0 \\ \tilde{\hat{K}} & 0 \end{bmatrix}\;.
\end{equation*}
Consequently, in \eqref{eq:projected_solution_operator_identity} the evolution of
the resolved moments is decoupled from the unresolved moments. However, the resolved
part of the memory kernel $\hat{\hat{K}}(t)$ involves the orthogonal dynamics
$e^{tR\mat{F}}$, whose solution is as complex as the full system \eqref{eq:linear_system}.
In the following, we derive new closures, by approximating the memory term in two
steps.
First, the integral in \eqref{eq:projected_solution_operator_identity} is
approximated using a quadrature rule. This is a general step that is independent of
the specific form of the differential operator $R$. We present various approximations
in Sect.~\ref{subsec:approx_memory_integral}.
Second, the exact orthogonal dynamics $e^{tR\mat{F}}$ are approximated by a simpler
evolution operator. Good approximations employ the specific structure of $R\mat{F}$.
We present such an approximation in Sect.~\ref{sec:apply_op_radiation}.

\subsection{Approximations of the Memory Integral}
\label{subsec:approx_memory_integral}
The simplest approximation of the memory integral is to drop it. This yields the
\emph{first order optimal prediction} system
\begin{equation}
\partial_t\vec{u} = \mathcal{R}\vec{u}\;,
\label{eq:linear_system_foop}
\end{equation}
which in some applications yields satisfying results. However, in most cases better
results can be expected when approximating the memory term more accurately. This
approach is generally denoted \emph{second order optimal prediction}. Assume for the
moment that the memory kernel $K(t)$ be exactly accessible. The memory term in
\eqref{eq:projected_solution_operator_identity} uses the solution at all previous times.
In principle, one could store the solution at all previous times, and thus
approximate the integral very accurately (as presented in \cite{Seibold2001}).
However, such approaches are typically very costly in computation time and memory.
In addition, for partial differential equations, the stability of such
integro-differential equations (or delay equations, after approximating the integral),
is a very delicate aspect. Therefore, here we focus on approximation that only use
the current state of the system. We present three approximations.
\begin{itemize}
\item\textbf{Piecewise constant quadrature:}
Assume the memory kernel decays with a characteristic time scale $\tau$.
Then a piecewise constant quadrature rule for $K(s)u(t-s)$ yields
\begin{equation}
\int_0^t K(s)u(t-s)\ud{s} \approx \int_0^\tau K(0)u(t)\ud{s} = \tau K(0)u(t)\;.
\label{eq:pw_const_quad}
\end{equation}
\item\textbf{Truncated piecewise constant quadrature:}
For $t<\tau$, the memory integral in \eqref{eq:pw_const_quad} cannot reach over the
full length $\tau$. Hence, it is reasonable to approximate
\begin{equation}
\int_0^t K(s)u(t-s)\ud{s} \approx \min\{\tau,t\} K(0)u(t)\;.
\label{eq:trunc_pw_const_quad}
\end{equation}
\item\textbf{Modified piecewise constant quadrature:}
Using a piecewise constant quadrature rule for $u(t-s)$ only, yields
\begin{equation}
\int_0^t K(s)u(t-s)\ud{s} \approx \int_0^t K(s)u(t)\ud{s} = \int_0^t K(s)\ud{s}\,u(t)\;.
\label{eq:mod_pw_const_quad}
\end{equation}
\end{itemize}

\section{Application to the Radiation Moment System}
\label{sec:apply_op_radiation}
We now turn our attention to the infinite moment system \eqref{eq:radiative_moment_system}.
For consistency with the notation used in Sect.~\ref{subsec:linear_optimal_prediction},
we denote the (infinite) vector of moments by
$\vec{u}(x,t) = \prn{u_0(x,t),u_1(x,t),\dots}^T$.
In addition, we neglect the forcing term, since it is unaffected by any truncation.
The radiation system \eqref{eq:radiative_moment_system} can be written as
\begin{equation}
\partial_t\vec{u} = R\vec{u}\;,
\label{eq:radiative_moment_system_sourcefree}
\end{equation}
where the differential operator $R = -\mat{B}\partial_x-\mat{C}$ involves the (infinite)
matrices \eqref{eq:radiative_moment_matrices}. We consider a splitting of the
moments $\vec{u} = \prn{\hat{\vec{u}},\tilde{\vec{u}}}$, where
$\hat{\vec{u}} = \prn{u_0,\dots}$ contains the moments resolved. The radiative
intensity $u_0$ is always the starting moment. The other moments, however, could be
reordered. The system splits into blocks
\begin{equation*}
\vec{u} = \begin{bmatrix} \hat{\vec{u}} \\ \tilde{\vec{u}} \end{bmatrix}
\; \text{,}\quad
\mat{B} = \begin{bmatrix}
\hat{\hat{\mat{B}}} & \hat{\tilde{\mat{B}}} \\
\tilde{\hat{\mat{B}}} & \tilde{\tilde{\mat{B}}}
\end{bmatrix}
\;, \text{and}\quad
\mat{C} = \begin{bmatrix}
\hat{\hat{\mat{C}}} & \mat{0} \\
\mat{0} & \tfrac{1}{\tau}\mat{I}
\end{bmatrix}\;,
\end{equation*}
where $\tau = \frac{1}{\kappa+\sigma}$, which we also assume to be the characteristic
time scale of the system ($\tau$ is a time scale since we have set $c=1$).
The optimal prediction system \eqref{eq:projected_solution_operator_identity}
involves the block operators
\begin{equation*}
R\mat{E}
= -\begin{bmatrix} \hat{\hat{\mat{B}}}\partial_x+\hat{\hat{\mat{C}}} & \mat{0} \\
\tilde{\hat{\mat{B}}}\partial_x & \mat{0} \end{bmatrix}
\; \text{,}\quad
R\mat{F}
= -\begin{bmatrix} \mat{0} & \hat{\tilde{\mat{B}}}\partial_x \\
\mat{0} & \tilde{\tilde{\mat{B}}}\partial_x+\tfrac{1}{\tau}\mat{I} \end{bmatrix}
\;, \text{and}\quad
R\mat{F}R\mat{E}
= \begin{bmatrix} \hat{\tilde{\mat{B}}}\tilde{\hat{\mat{B}}}\partial_{xx} & \mat{0} \\
\tilde{\tilde{\mat{B}}}\tilde{\hat{\mat{B}}}\partial_{xx}
+\tfrac{1}{\tau}\tilde{\hat{\mat{B}}}\partial_{x} & \mat{0} \end{bmatrix}\;.
\end{equation*}
Here the upper left block of $R\mat{E}$ is the projected right hand side, which
yields the first order optimal prediction system. The operator $R\mat{F}$ describes
the orthogonal dynamics system
\begin{equation}
\begin{cases}
\partial_t\hat{\vec{u}}   &= -\hat{\tilde{\mat{B}}}\partial_x\tilde{\vec{u}} \\
\partial_t\tilde{\vec{u}} &= -\tilde{\tilde{\mat{B}}}\partial_x\tilde{\vec{u}}
-\tfrac{1}{\tau}\tilde{\vec{u}}\;.
\end{cases}
\label{eq:orthogonal_dynamics}
\end{equation}
Its solution operator $e^{tR\mat{F}}$ is contained in the memory
kernel \eqref{eq:memory_kernel}. The equation for $\tilde{\vec{u}}$
in \eqref{eq:orthogonal_dynamics} is similar to the original full
system \eqref{eq:radiative_moment_system_sourcefree}, however, with one fundamental
simplification. The decay matrix is a multiple of the identity, a fact that may be
employed for fast approximations of the orthogonal dynamics. We shall pursue this approach
in future work. For now, we construct a simple approximate system by neglecting the
advection in the unresolved variables, i.e.~$\tilde{\tilde{\mat{B}}} \approx \mat{0}$.
The approximate orthogonal dynamics operator
\begin{equation*}
G = -\begin{bmatrix} \mat{0} & \hat{\tilde{\mat{B}}}\partial_x \\
\mat{0} & \tfrac{1}{\tau}\mat{I} \end{bmatrix}
\approx R\mat{F}
\end{equation*}
has a simple solution
\begin{equation*}
e^{tG}\begin{bmatrix} \hat{\vec{u}} \\ \tilde{\vec{u}} \end{bmatrix}
= \begin{bmatrix} \hat{\vec{u}}-\tau\prn{1-e^{-t/\tau}}
\hat{\tilde{\mat{B}}}\partial_x\tilde{\vec{u}} \\
e^{-t/\tau}\tilde{\vec{u}}
\end{bmatrix}\;.
\end{equation*}
Hence, the resolved component of the approximate memory kernel is given by
\begin{align*}
\widehat{\widehat{e^{tG}R\mat{F}R\mat{E}}}
&= \hat{\tilde{\mat{B}}}\tilde{\hat{\mat{B}}}\partial_{xx}
-\tau\prn{1-e^{-t/\tau}}\hat{\tilde{\mat{B}}}\partial_x
\prn{\tilde{\tilde{\mat{B}}}\tilde{\hat{\mat{B}}}\partial_{xx}
+\tfrac{1}{\tau}\tilde{\hat{\mat{B}}}\partial_{x}} \\
&= e^{-t/\tau}\hat{\tilde{\mat{B}}}\tilde{\hat{\mat{B}}}\partial_{xx}
-\tau\prn{1-e^{-t/\tau}}\hat{\tilde{\mat{B}}}\tilde{\tilde{\mat{B}}}
\tilde{\hat{\mat{B}}}\partial_{xxx}\;.
\end{align*}
In the following, we work out the specific expressions for two different types
of ordering of the moments.

\subsection{Standard ordering}
\label{subsec:apply_op_radiation_standard}
We consider $\hat{\vec{u}} = \prn{u_0,u_1,\dots,u_N}$ and
$\tilde{\vec{u}} = \prn{u_{N+1},u_{N+2},\dots}$.
With respect to this ordering, the projected right hand side
$\hat{\hat{\mathcal{R}}}$ is a simple truncated version of the full right hand
side. The triple matrix product vanishes
$\hat{\tilde{\mat{B}}}\tilde{\tilde{\mat{B}}}\tilde{\hat{\mat{B}}} = \mat{0}$.
And the double matrix product equals
\begin{equation*}
\hat{\tilde{\mat{B}}}\tilde{\hat{\mat{B}}}
= \begin{pmatrix}
0 & \hdots & 0 \\
\vdots & \ddots & \vdots \\
0 & \hdots & \vartheta_N
\end{pmatrix}\;,
\end{equation*}
where $\vartheta_N = \frac{(N+1)^2}{(2N+1)(2N+3)}$.
Hence, with the above simplified orthogonal dynamics, the optimal prediction
integro-differential equation \eqref{eq:projected_solution_operator_identity} equals
\begin{equation}
\partial_t\hat{\vec{u}}(t) = -\hat{\hat{\mat{B}}}\partial_x\hat{\vec{u}}(t)
-\hat{\hat{\mat{C}}}\hat{\vec{u}}(t)
+\int_0^t e^{-s/\tau}\hat{\tilde{\mat{B}}}\tilde{\hat{\mat{B}}}\partial_{xx}
\hat{\vec{u}}(t-s)\ud{s}\;.
\label{eq:op_system_standard}
\end{equation}
Due to the special structure of $\hat{\tilde{\mat{B}}}\tilde{\hat{\mat{B}}}$,
the memory term only modifies the $N$-th equation, thus it is in fact a closure.
The various approximations to the memory integral, presented in
Sect.~\ref{subsec:approx_memory_integral}, yield the following modifications
to the equation of the $N$-th moment:
\begin{itemize}
\item\textbf{First order optimal prediction:}
When dropping the integral term, the resulting system \eqref{eq:op_system_standard}
becomes exactly the $P_N$ closure. Note that by considering measures with
non-vanishing covariances between the moments, all other linear closures can be
derived \cite{FrankSeibold2008}.
\item\textbf{Piecewise constant quadrature:}
The integral term in \eqref{eq:op_system_standard} is approximated by
\begin{equation}
\tau\vartheta_N\partial_{xx}u_N(t)\;.
\label{eq:int_approx_diffusion}
\end{equation}
A comparison with \eqref{eq:closure_diffusion_correction} reveals that we have
re-derived Levermore's diffusion correction closure \cite{Levermore2005}.
In the case $N=0$, it is equivalent to the classical diffusion approximation.
\item\textbf{Truncated/modified piecewise constant quadrature:}
The approximations \eqref{eq:trunc_pw_const_quad}, respectively
\eqref{eq:mod_pw_const_quad} lead to the integral approximation
\begin{equation}
f(t)\vartheta_N\partial_{xx}u_N(t)\;.
\label{eq:int_approx_crescendo_diffusion}
\end{equation}
Compared to the diffusion correction closure \eqref{eq:int_approx_diffusion},
the constant $\tau$ is replaced by a time dependent function $f(t)$, that increases
gradually towards its maximum value $\tau$. The diffusion is turned on gradually over
time. Hence, we denote these types of approximations \emph{crescendo diffusion} ($N=0$),
respectively \emph{crescendo diffusion correction} ($N>0$) closure.
More specifically, we call the approximation \emph{truncated crescendo diffusion (closure)}
when it is based on \eqref{eq:trunc_pw_const_quad}. In this case $f(t) = \min\{\tau,t\}$.
Similarly, the approximation based on \eqref{eq:mod_pw_const_quad}, that leads to
$f(t) = \tau\prn{1-e^{-t/\tau}}$, is called \emph{modified crescendo diffusion (closure)}.
\end{itemize}
The new crescendo closures \eqref{eq:int_approx_crescendo_diffusion} introduce an
explicit time dependence in the diffusion term. The physical rationale is
that at $t=0$, the state of the system (at least the unknowns of interest)
is known exactly. Information is lost as time evolves, due to the approximation.
This implies that crescendo diffusion closures shall only be used if the initial
conditions are known with higher accuracy than the approximation error due to
truncation. It is one way of circumventing an initial layer in time.
In Sect.~\ref{subsec:numerical_results_diffusion}, we investigate the quality of these
new approximations numerically.

Although here we have assumed spatially homogeneous coefficients, we expect that
the equations can be adapted to the space-dependent case in analogy to diffusion theory.
Specifically, if $\kappa(x)$ and $\sigma(x)$ are space dependent, we define
$\tau(x) = \frac{1}{\kappa(x)+\sigma(x)}$, and replace
$\tau\partial_{xx}\hat{u}_N$ by $\partial_x\prn{\tau(x)\partial_x\hat{u}_N}$.
The validity of this approximation will be addressed in future research.

\subsection{Reordered equations}
\label{subsec:apply_op_radiation_reordered}
In the following, we consider a reordering of the unknowns, given by the reordered vector
$\vec{v} = \mat{P}\vec{u}$, where $\mat{P}$ is a permutation matrix that leaves $v_0 = u_0$
as the first component. This reordering is inspired by Gelbard's original
derivation \cite{Gelbard1960,Gelbard1961,Gelbard1962} of the steady simplified $P_N$
equations, in which the unknowns are the even-order moments and the odd-order moments are
algebraically eliminated. The reordered system is
\begin{equation*}
\partial_t \vec{v} = -\mathcal{B}\partial_x\vec{v}-\mat{C}\vec{v}\;,
\end{equation*}
where we have the permuted advection matrix $\mathcal{B} = \mat{P}\mat{B}\mat{P}^T$.
The decay term remains unchanged, since $\mat{P}\mat{C}\mat{P}^T = \mat{C}$.
Of particular interest is the following \emph{even-odd reordering}:
For a given number $N$, consider the following permutation of the unknowns
\begin{equation*}
\vec{v} = \mat{P}\vec{u}
= \prn{u_0,u_2,\dots,u_{2N},u_1,u_3,\dots,u_{2N+1},u_{2N+2},\dots}^T\;,
\end{equation*}
and consider the $N+1$ lowest even moments $\hat{\vec{v}} = \prn{u_0,u_2,\dots,u_{2N}}$
as of interest. The resulting reordered advection matrix has the following structure
(here for $N=2$):
\begin{equation*}
\begin{bmatrix} \hat{\hat{\mathcal{B}}} & \hat{\tilde{\mathcal{B}}} \\
\tilde{\hat{\mathcal{B}}} & \tilde{\tilde{\mathcal{B}}} \end{bmatrix}
= \left[\begin{array}{ccc|cccccc}
 & & & 1 & & & & & \\
 & & & 2/5 & 3/5 & & & & \\
 & & &  & 4/9 & 5/9 & & & \\
\hline
1/3 & 2/3 & & & & & & & \\
 & 3/7 & 4/7 & & & & & & \\
 & & 5/11 & & & & 6/11 & & \\
 & & & & & 6/13 & & 7/13 & \\
 & & & & & & 7/15 & & \ddots \\
 & & & & & & & \ddots &
\end{array}\right]\;.
\end{equation*}
The solid lines indicate separations between $\hat{\vec{v}}$ and $\tilde{\vec{v}}$.
For the even-odd reordering, we have $\hat{\hat{\mathcal{B}}} = \mat{0}$.
As for standard ordering, the triple matrix product vanishes
$\hat{\tilde{\mathcal{B}}}\tilde{\tilde{\mathcal{B}}}\tilde{\hat{\mathcal{B}}} = \mat{0}$.
The double matrix product $\mat{D} = \hat{\tilde{\mathcal{B}}}\tilde{\hat{\mathcal{B}}}$
becomes more interesting than for standard ordering. For any choice of $N$, it is observed
to be positive definite. In addition, lower order matrices are contained as submatrices
in higher-order ones. With the simplified orthogonal dynamics, the optimal prediction
integro-differential equation \eqref{eq:projected_solution_operator_identity} equals
\begin{equation}
\partial_t\hat{\vec{v}}(t) = -\hat{\hat{\mat{C}}}\hat{\vec{v}}(t)
+\int_0^t e^{-s/\tau}\mat{D}\partial_{xx}\hat{\vec{v}}(t-s)\ud{s}\;.
\label{eq:op_system_reordered}
\end{equation}
The various approximations to the memory integral, presented in
Sect.~\ref{subsec:approx_memory_integral}, now become:
\begin{itemize}
\item\textbf{First order optimal prediction:}
Dropping the integral term in \eqref{eq:op_system_reordered} yields
a mere exponential decay for each moment
\begin{equation*}
\partial_t\hat{\vec{v}} = -\hat{\hat{\mat{C}}}\hat{\vec{v}}\;,
\end{equation*}
which is obviously a horrible approximation. For even-odd reordering, a non-trivial
approximation to the integral term is strictly required.
\item\textbf{Piecewise constant quadratures:}
System \eqref{eq:op_system_reordered} is approximated by
\begin{equation}
\partial_t\hat{\vec{v}} = -\hat{\hat{\mat{C}}}\hat{\vec{v}}
+f(t)\mat{D}\partial_{xx}\hat{\vec{v}}\;,
\label{eq:RPN}
\end{equation}
where the coefficient function $f(t)$ may be constant or time-dependent, depending
on the integral approximation used. When based on \eqref{eq:pw_const_quad}, we have
$f(t) = \tau$. In this case, we denote system \eqref{eq:RPN}
the \emph{reordered $P_N$ equations} ($RP_N$).
When the approximation \eqref{eq:trunc_pw_const_quad} is used, we have
$f(t) = \min\{\tau,t\}$, and speak of the \emph{truncated crescendo $RP_N$ equations}.
Similarly, when \eqref{eq:mod_pw_const_quad}, we have $f(t) = \tau\prn{1-e^{-t/\tau}}$,
and speak of the \emph{modified crescendo $RP_N$ equations}.
In the case $N=0$, these systems become the corresponding
diffusion closures \eqref{eq:closure_diffusion}.
\end{itemize}
The $RP_N$ system \eqref{eq:RPN} has strong similarity with the time dependent
$SP_N$ equations \cite{FrankKlarLarsenYasuda2007}. These are derived via asymptotic
analysis. Although the unknowns are not directly comparable, a comparison of the
respective diffusion matrices is compelling. For the choice of $\alpha=\tfrac{1}{3}$
the $SP_3$ diffusion matrix is
\begin{equation*}
\mat{D} = \begin{pmatrix}
    0.3333  &  0.6667  & -0.3333 \\
    0.1333  &  0.5238  &       0 \\
    0.3333  &  0.6667  &  0.2000
\end{pmatrix}\;.
\end{equation*}
In comparison, the $RP_2$ matrix is
\begin{equation*}
\mat{D} = \begin{pmatrix}
    0.3333  &  0.6667  &       0 \\
    0.1333  &  0.5238  &  0.3429 \\
         0  &  0.1905  &  0.5065
\end{pmatrix}\;.
\end{equation*}
Observe that the respective upper $2\times 2$ submatrices are identical.
The other entries differ, which can be expected, since the third unknown in the
$SP_3$ equations is not the fourth moment, as it is for the $RP_2$ equations.
Assuming the third variable in $SP_3$ be quasi-steady leads to the $SSP_3$ system
\cite{FrankKlarLarsenYasuda2007}, that turns out to be identical to
the $RP_1$ system. Hence, similarly as both classical and new diffusion approximations
can be derived from optimal prediction with standard ordering, even-odd ordering
allows the derivation of existing and new parabolic systems that approximate the
radiative transfer equations.
In Sect.~\ref{subsec:numerical_results_parabolic}, we investigate the quality of these
new systems numerically.

\section{Numerical Results}
\label{sec:numerical_results}
As a test case, we consider the 1D slab geometry
system \eqref{eq:radiation_equation_simple} on $x\in]0,1[$ (periodic boundary conditions)
with $\kappa = \sigma = 1.5$, and no source $q=0$. The initial conditions are
$u_0(x,0) = \exp\prn{-500\prn{x-\frac{1}{2}}^2}$ and $u_k(x,0) = 0\ \forall k\ge 1$.
The numerical solution is obtained using a second order staggered grid finite
difference method with resolution $\Delta x = 10^{-3}$ and $\Delta t = 0.8\Delta x$.
Any diffusive terms are implemented by two Crank-Nicolson half steps.

\begin{figure}[ht]
\centering
\begin{minipage}[t]{.49\textwidth}
\includegraphics[width=0.99\textwidth]{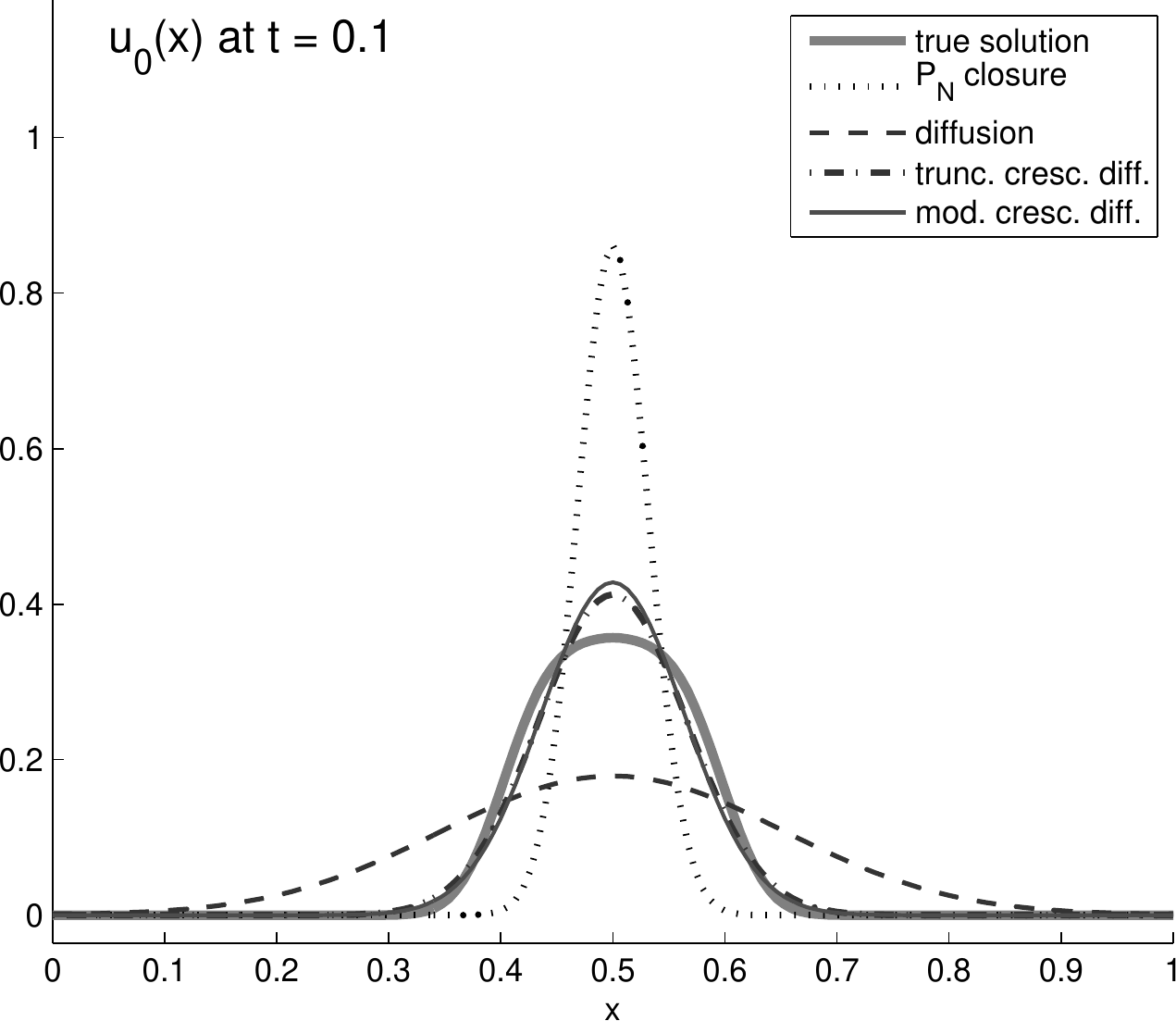}
\end{minipage}
\hfill
\begin{minipage}[t]{.49\textwidth}
\includegraphics[width=0.99\textwidth]{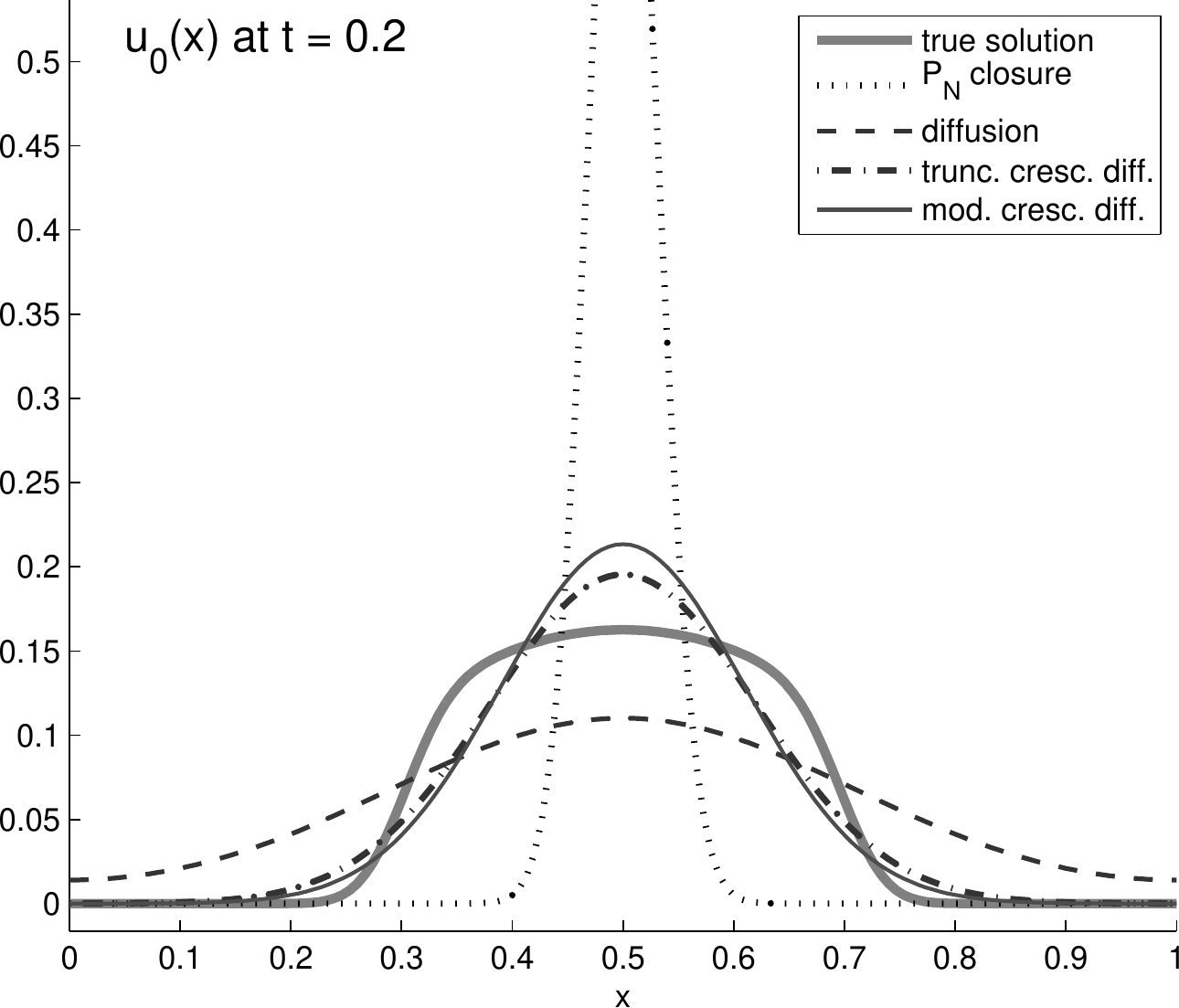}
\end{minipage}
\begin{minipage}[t]{.49\textwidth}
\includegraphics[width=0.99\textwidth]{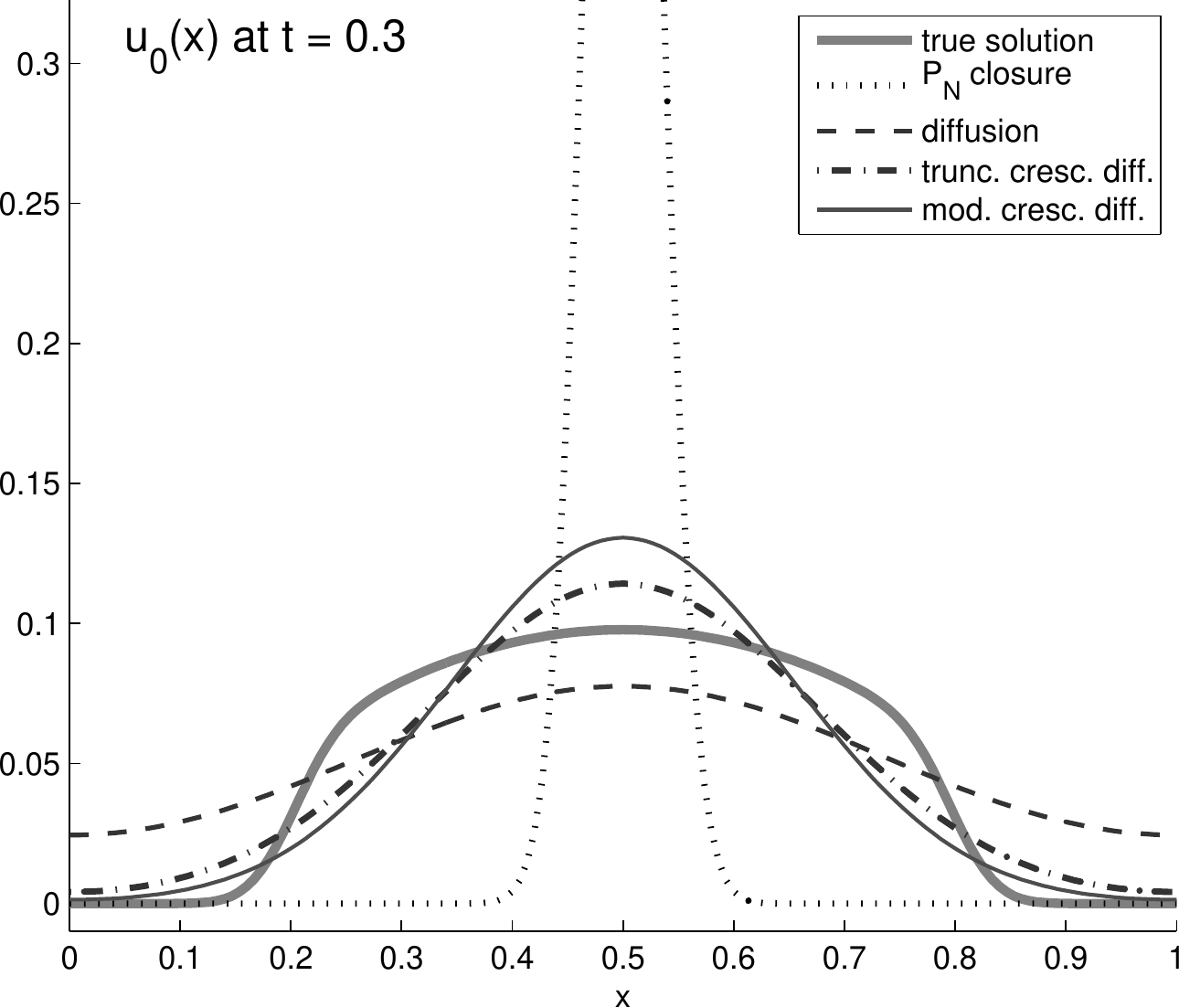}
\end{minipage}
\hfill
\begin{minipage}[t]{.49\textwidth}
\includegraphics[width=0.99\textwidth]{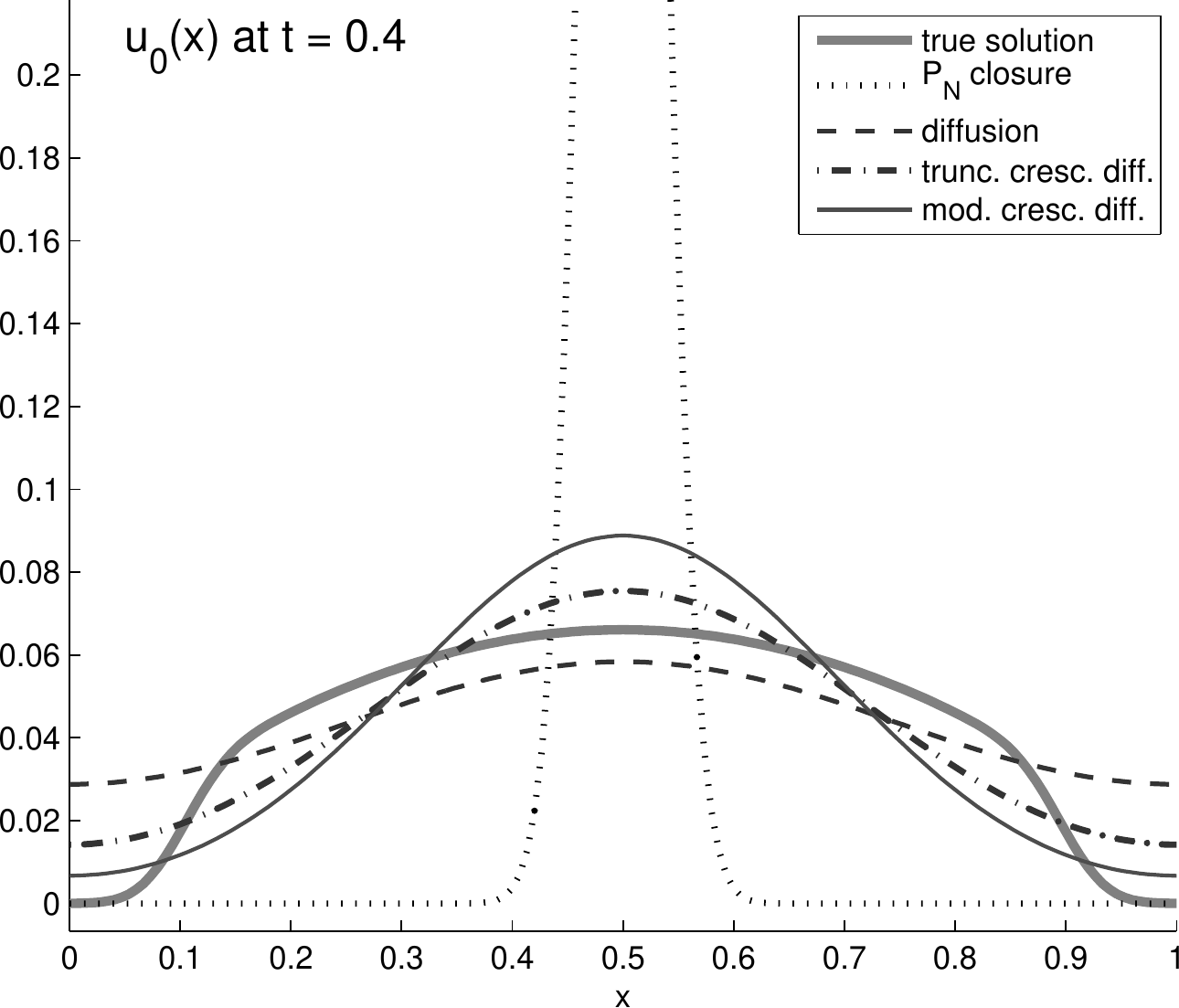}
\end{minipage}
\caption{Evolution of energy distribution for various moment closures for $N=0$}
\label{fig:radiation_1d_N0}
\end{figure}

\begin{figure}[p]
\centering
\begin{minipage}[t]{.49\textwidth}
\includegraphics[width=0.99\textwidth]{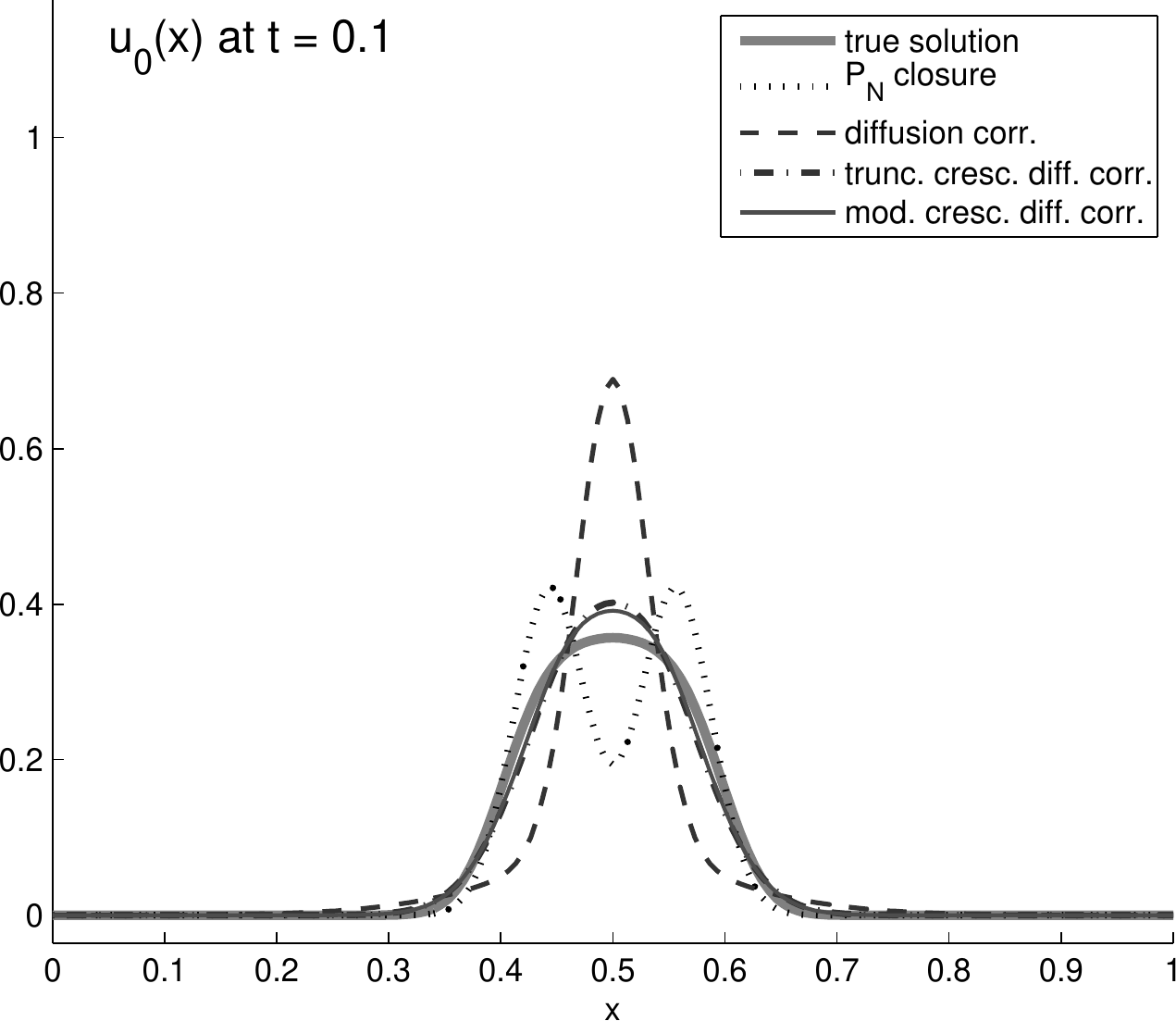}
\end{minipage}
\hfill
\begin{minipage}[t]{.49\textwidth}
\includegraphics[width=0.99\textwidth]{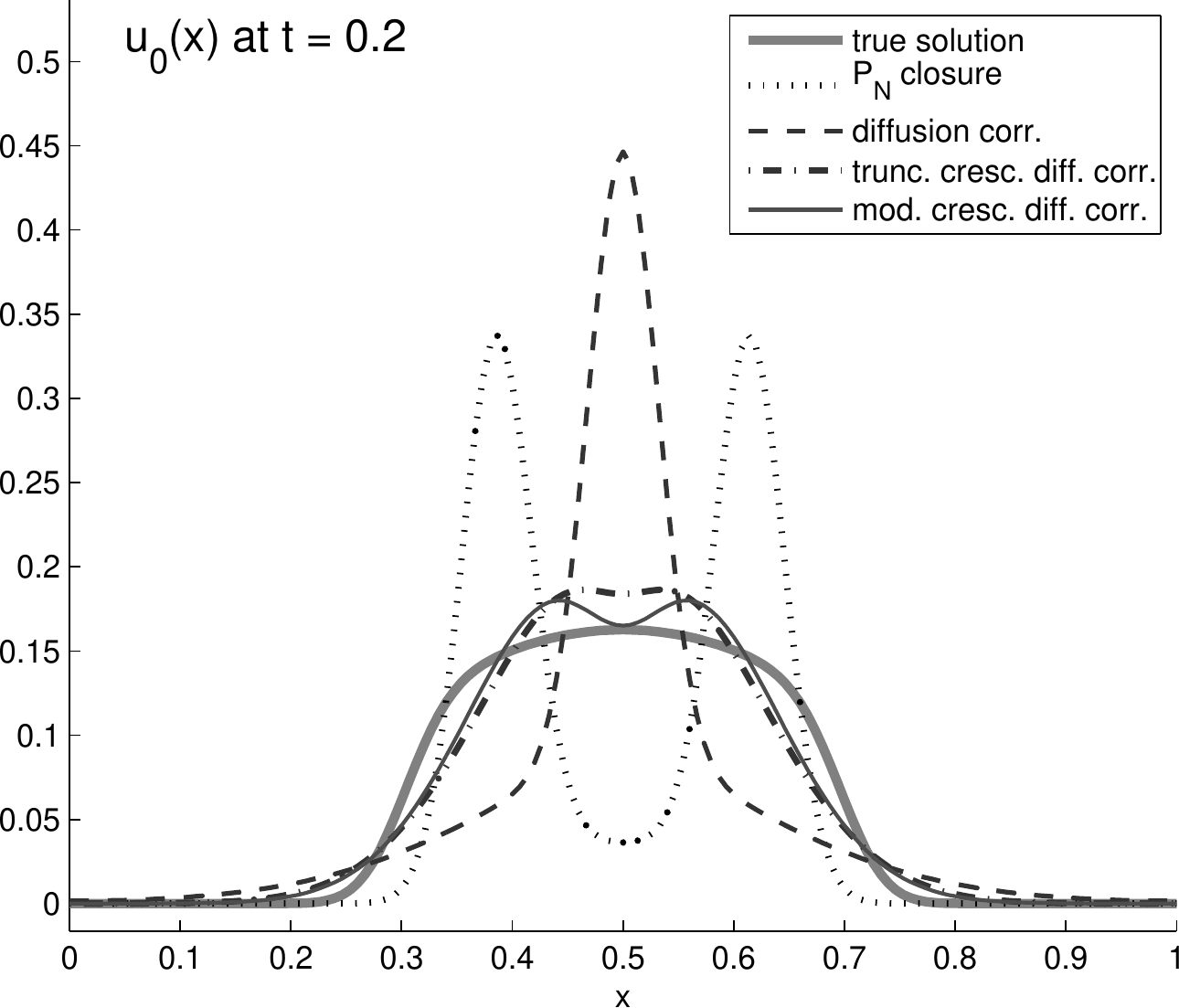}
\end{minipage}
\begin{minipage}[t]{.49\textwidth}
\includegraphics[width=0.99\textwidth]{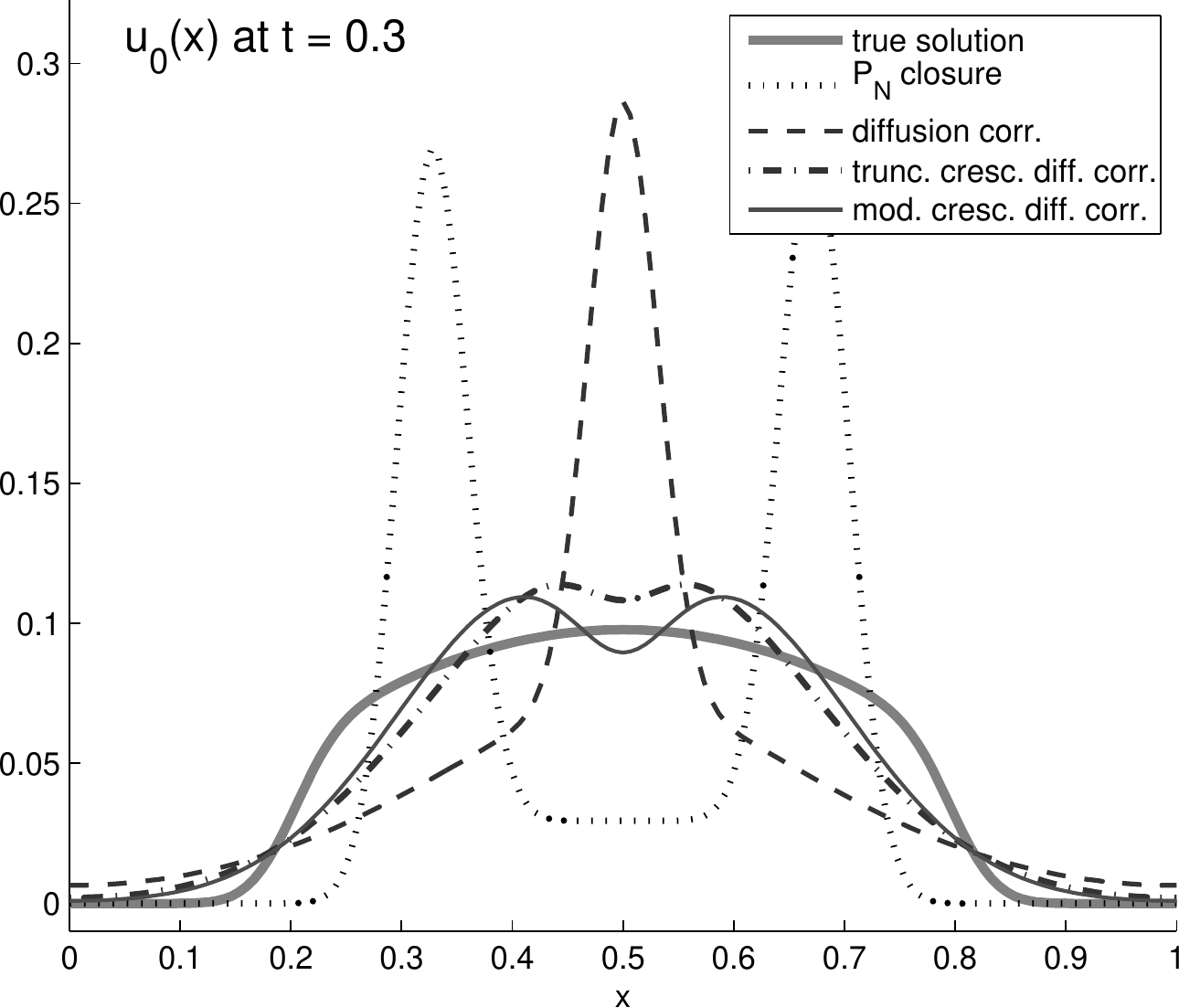}
\end{minipage}
\hfill
\begin{minipage}[t]{.49\textwidth}
\includegraphics[width=0.99\textwidth]{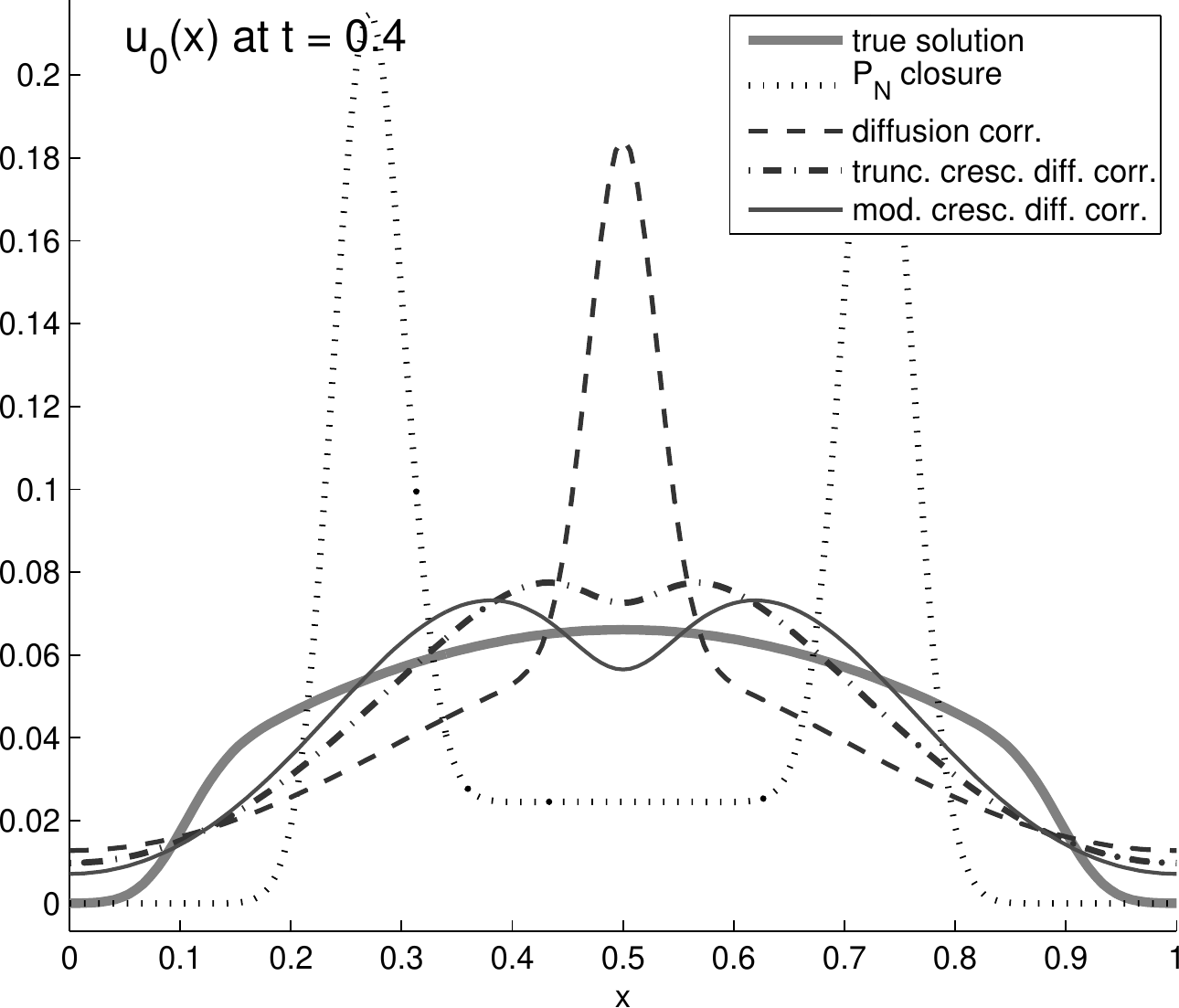}
\end{minipage}
\caption{Evolution of energy distribution for various moment closures for $N=1$}
\label{fig:radiation_1d_N1}
\vspace{2em}
\centering
\begin{minipage}[t]{.49\textwidth}
\includegraphics[width=0.99\textwidth]{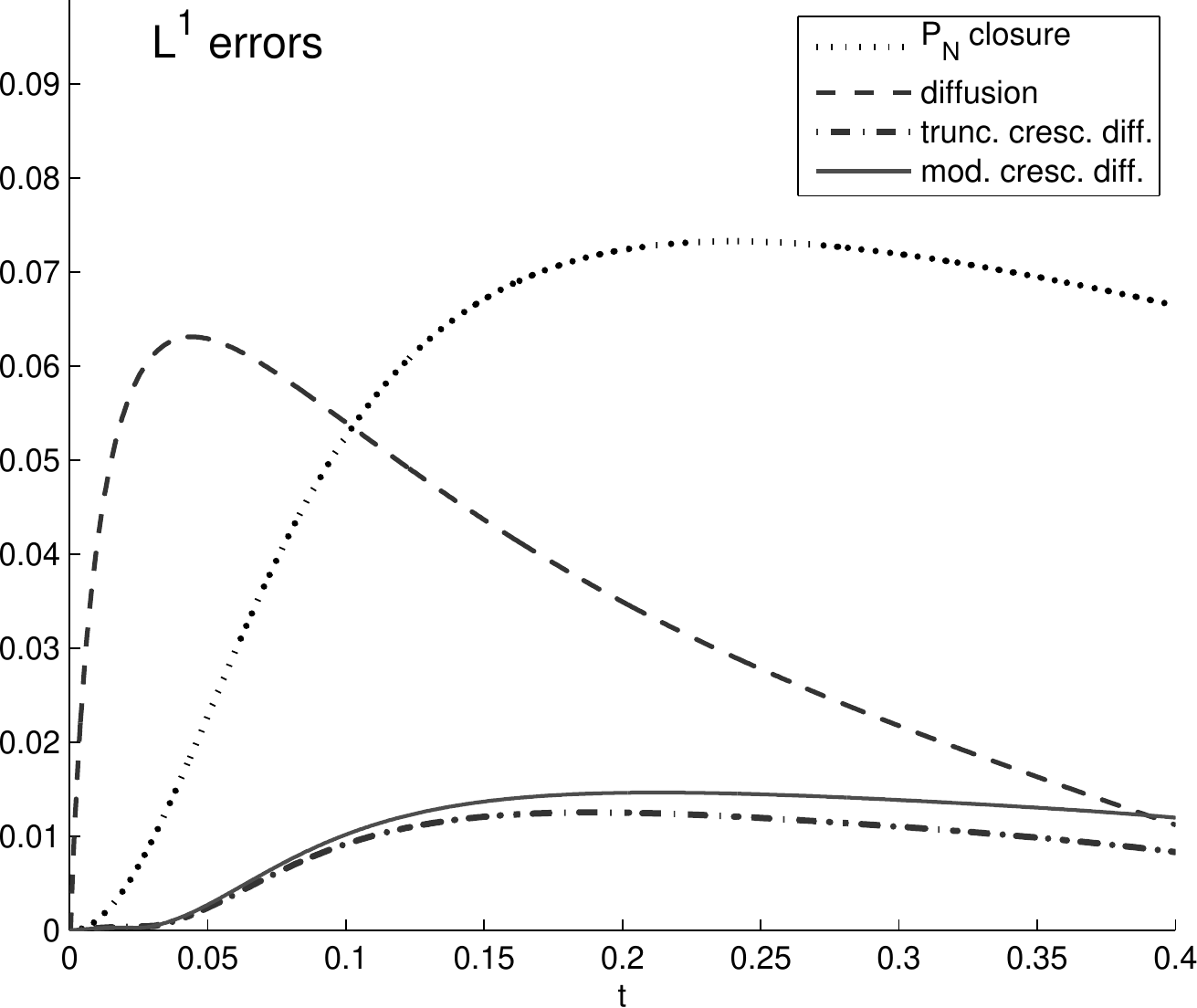}
\caption{Evolution of the $L^1$ error for moment closures for $N=0$}
\label{fig:radiation_1d_N0_error}
\end{minipage}
\hfill
\begin{minipage}[t]{.49\textwidth}
\includegraphics[width=0.99\textwidth]{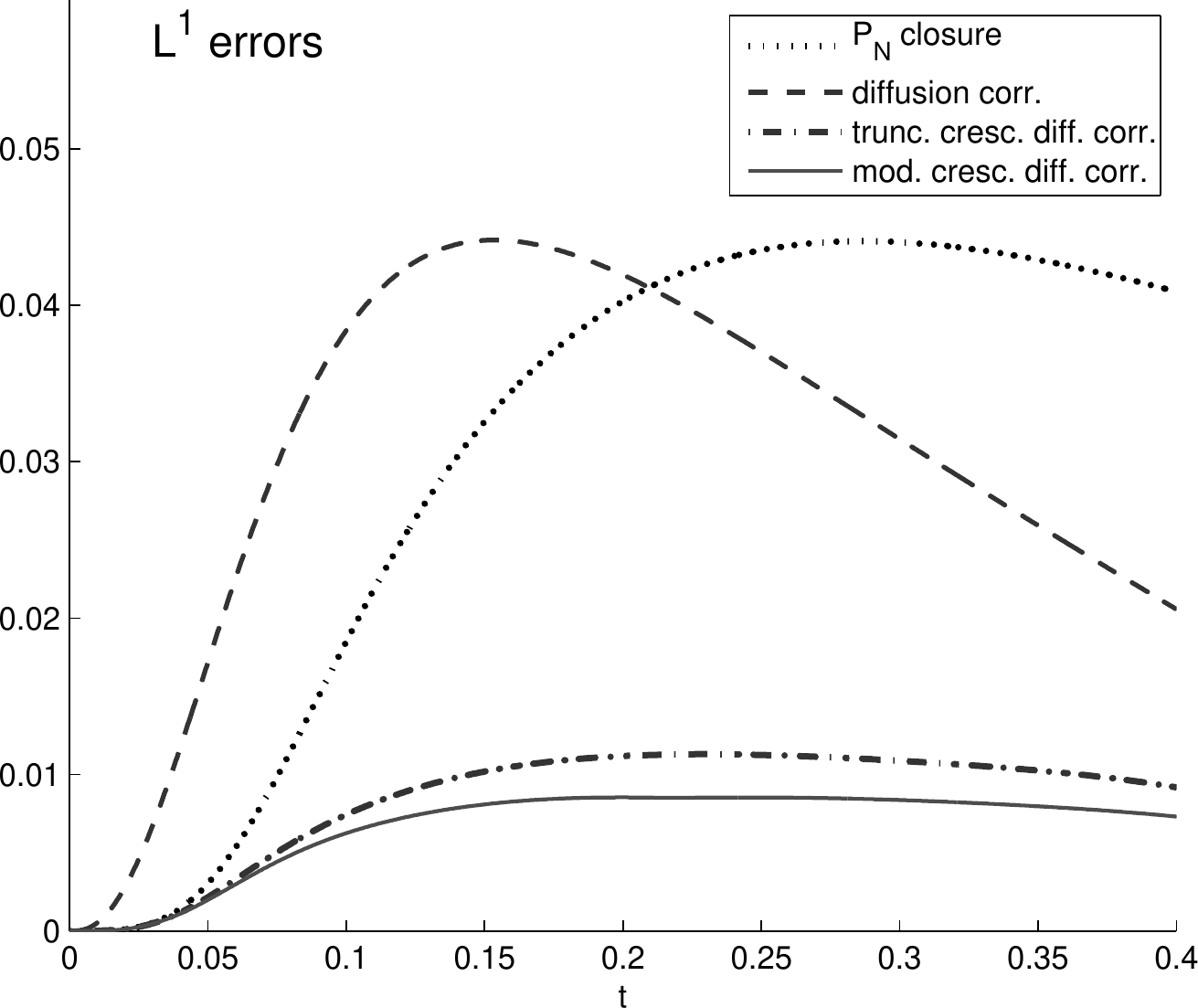}
\caption{Evolution of the $L^1$ error for moment closures for $N=1$}
\label{fig:radiation_1d_N1_error}
\end{minipage}
\end{figure}

\subsection{Diffusion Correction Approximations}
\label{subsec:numerical_results_diffusion}
\begin{figure}[ht]
\centering
\begin{minipage}[t]{.49\textwidth}
\includegraphics[width=0.99\textwidth]{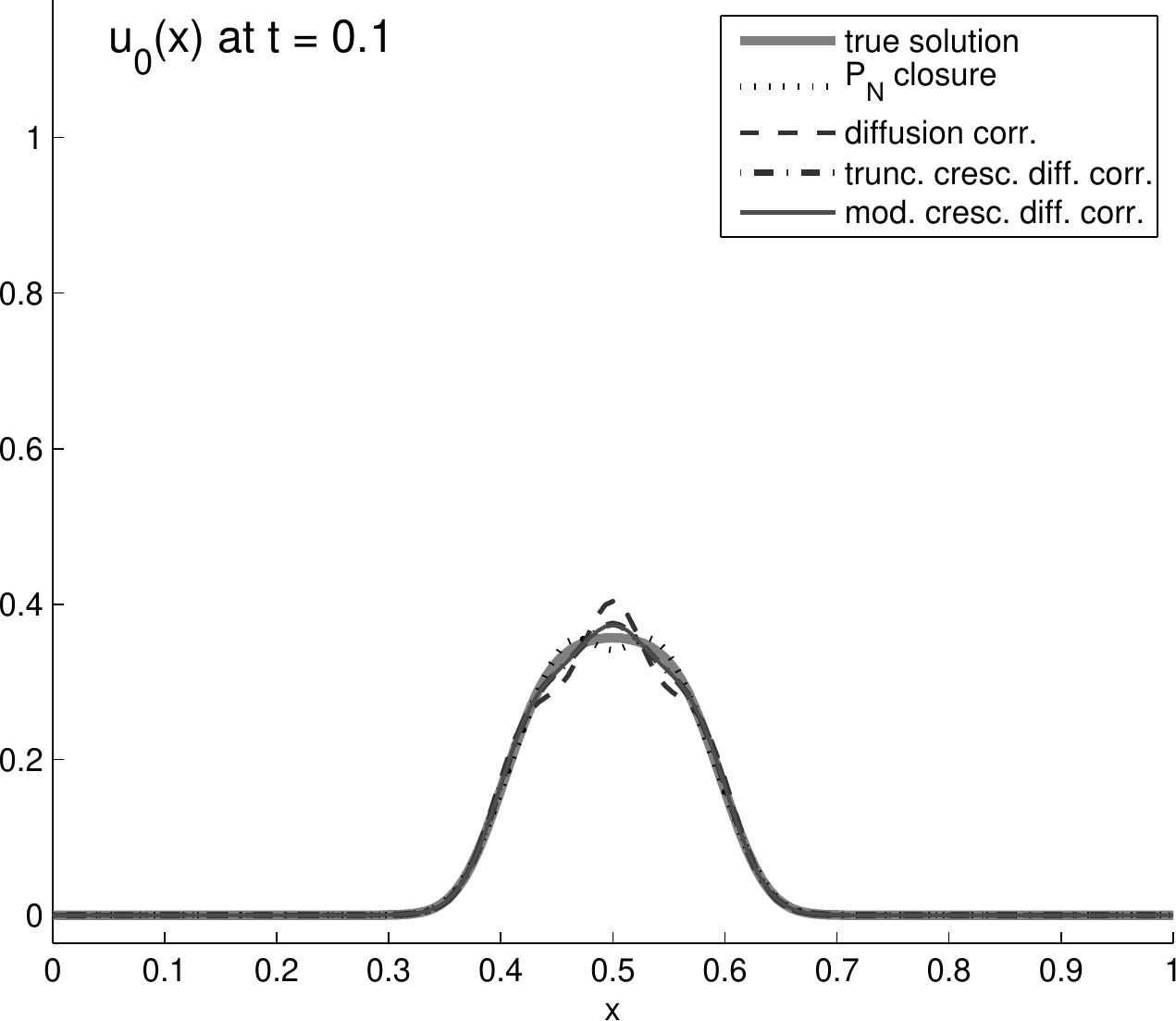}
\end{minipage}
\hfill
\begin{minipage}[t]{.49\textwidth}
\includegraphics[width=0.99\textwidth]{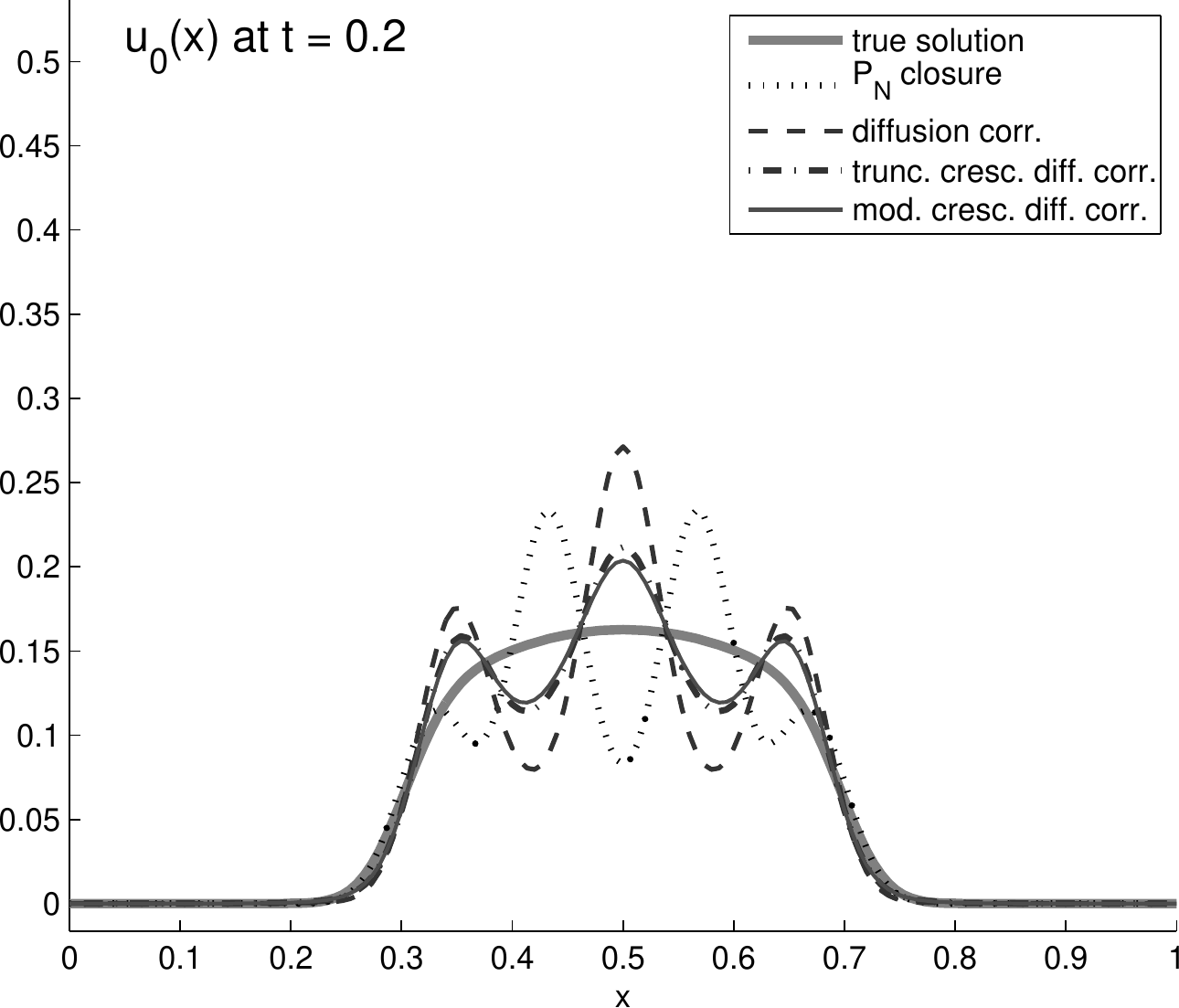}
\end{minipage}
\begin{minipage}[t]{.49\textwidth}
\includegraphics[width=0.99\textwidth]{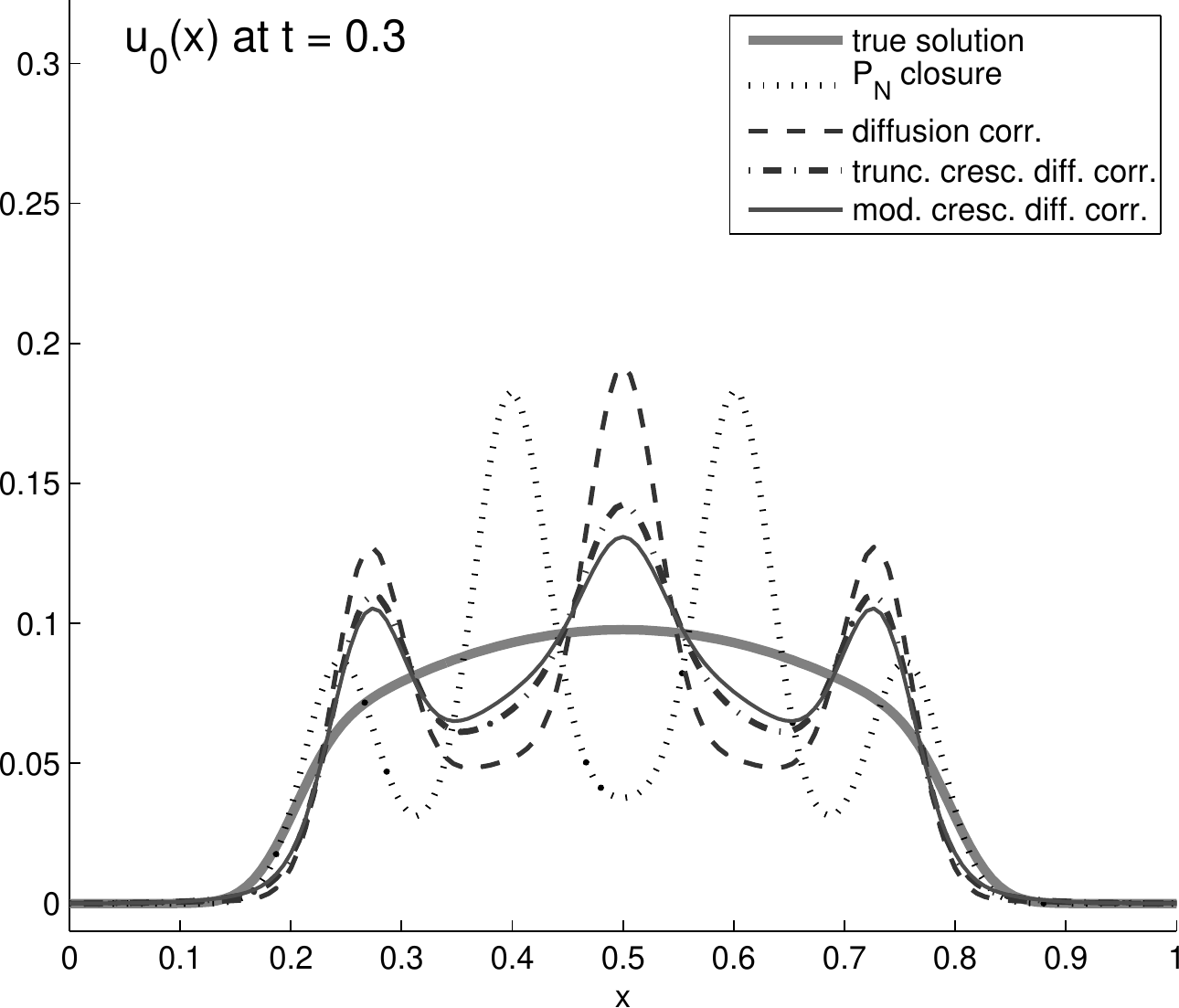}
\end{minipage}
\hfill
\begin{minipage}[t]{.49\textwidth}
\includegraphics[width=0.99\textwidth]{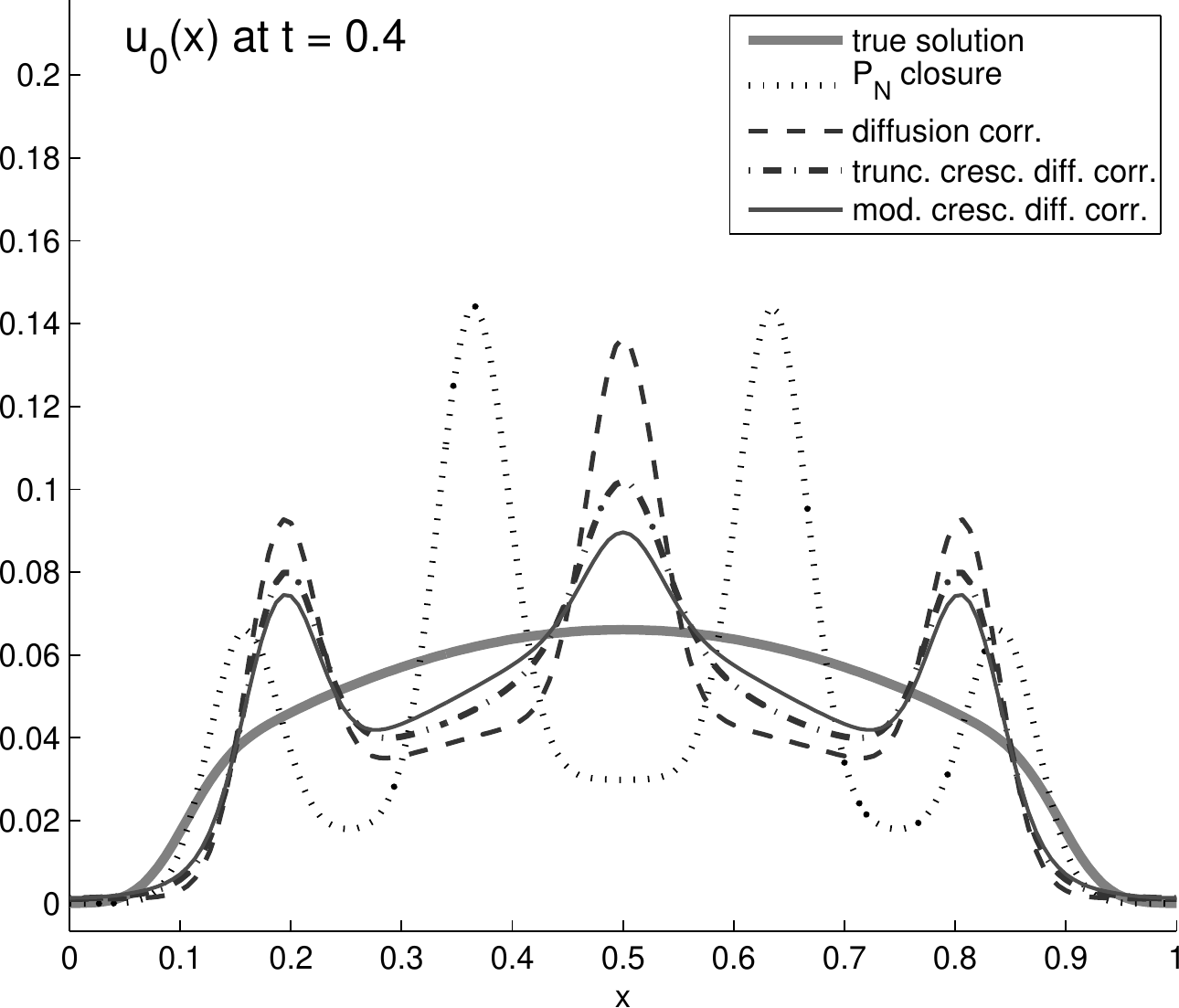}
\end{minipage}
\caption{Evolution of energy distribution for various moment closures for $N=3$}
\label{fig:radiation_1d_N3}
\end{figure}

The time evolution $t\in\{0.1,0.2,0.3,0.4\}$ of the radiative intensity $u_0(x,t)$ for
various system sizes is shown in Fig.~\ref{fig:radiation_1d_N0} for $N=0$,
Fig.~\ref{fig:radiation_1d_N1} for $N=1$, and Fig.~\ref{fig:radiation_1d_N3} for $N=3$.
In each case, the thick solid curve is the solution to the full radiative transfer equation,
obtained using a $P_N$ closure, which has converged in $N$ ($N=51$). The dotted function
is the classical $P_N$
closure (Sect.~\ref{subsec:closure_PN}). The dashed curve is the diffusion correction
closure \eqref{eq:closure_diffusion_correction}, which for $N=0$ equals the diffusion
closure \eqref{eq:closure_diffusion}. The dash-dotted curve and the thin solid curve are
the two versions of the
crescendo diffusion correction \eqref{eq:int_approx_crescendo_diffusion}, as derived
in Sect.~\ref{subsec:apply_op_radiation_standard}. The dash-dotted curve denotes the
sharp time dependence, while the solid curve denotes the exponential time dependence.

\begin{figure}[p]
\centering
\begin{minipage}[t]{.49\textwidth}
\includegraphics[width=0.99\textwidth]{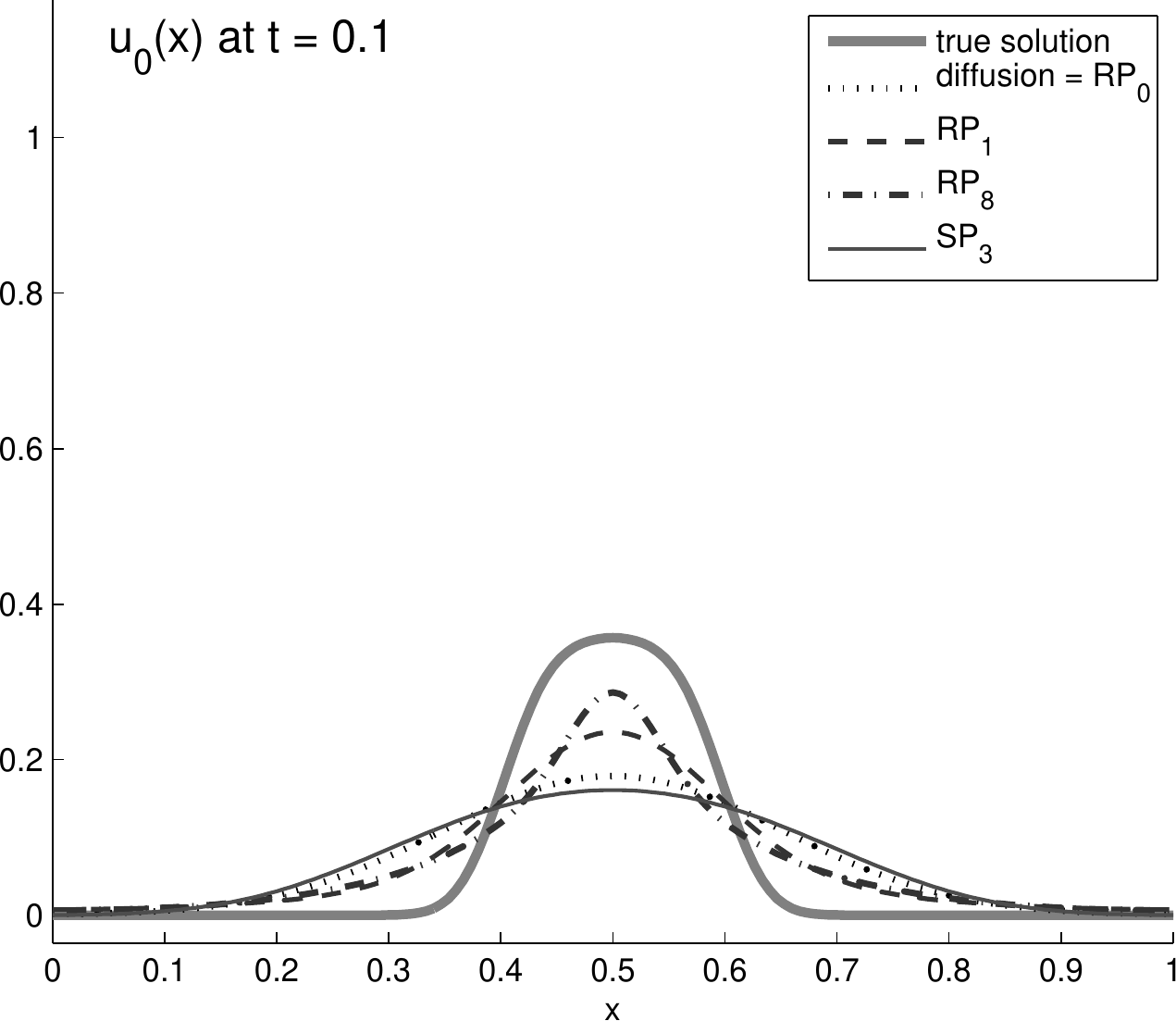}
\end{minipage}
\hfill
\begin{minipage}[t]{.49\textwidth}
\includegraphics[width=0.99\textwidth]{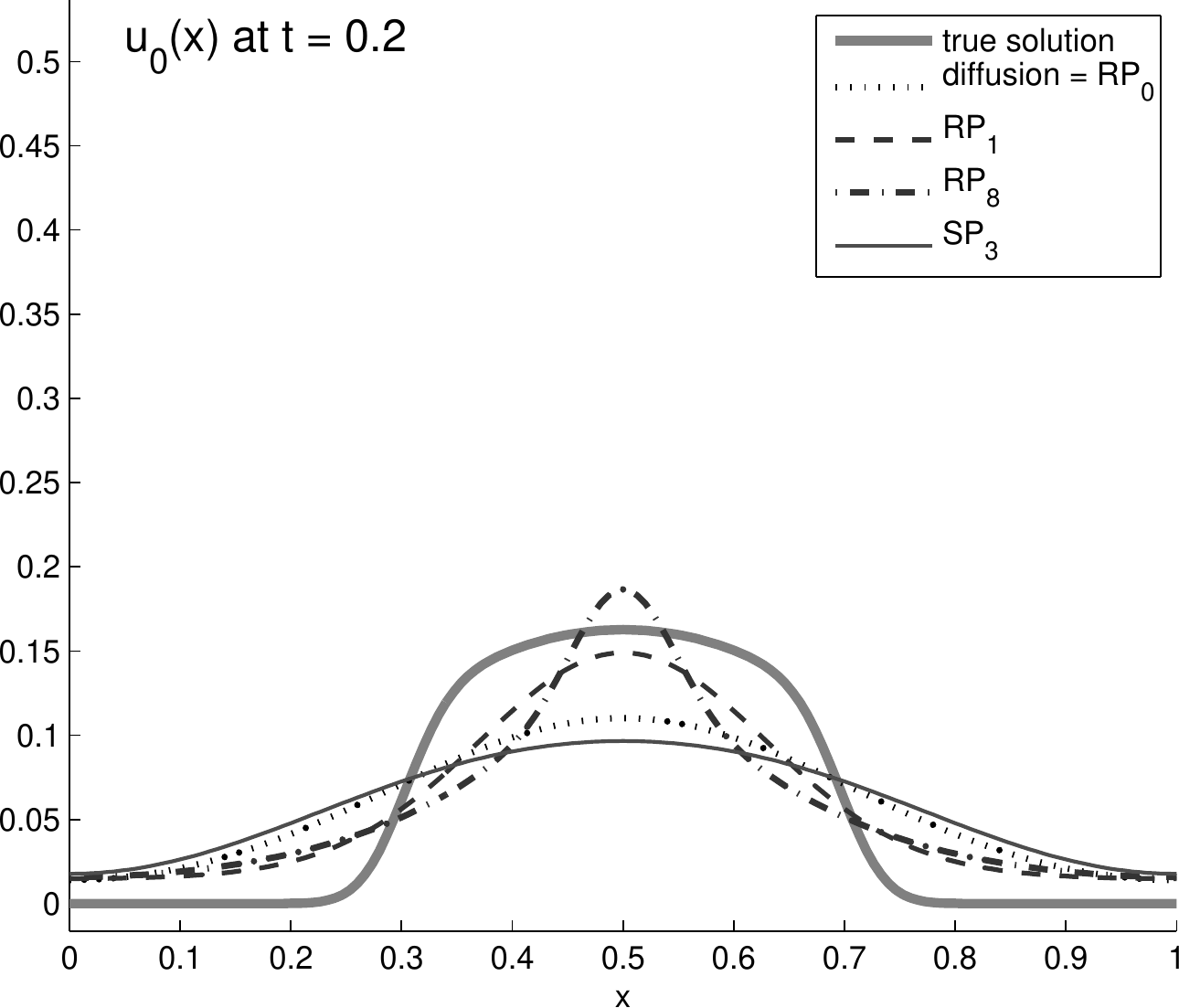}
\end{minipage}
\begin{minipage}[t]{.49\textwidth}
\includegraphics[width=0.99\textwidth]{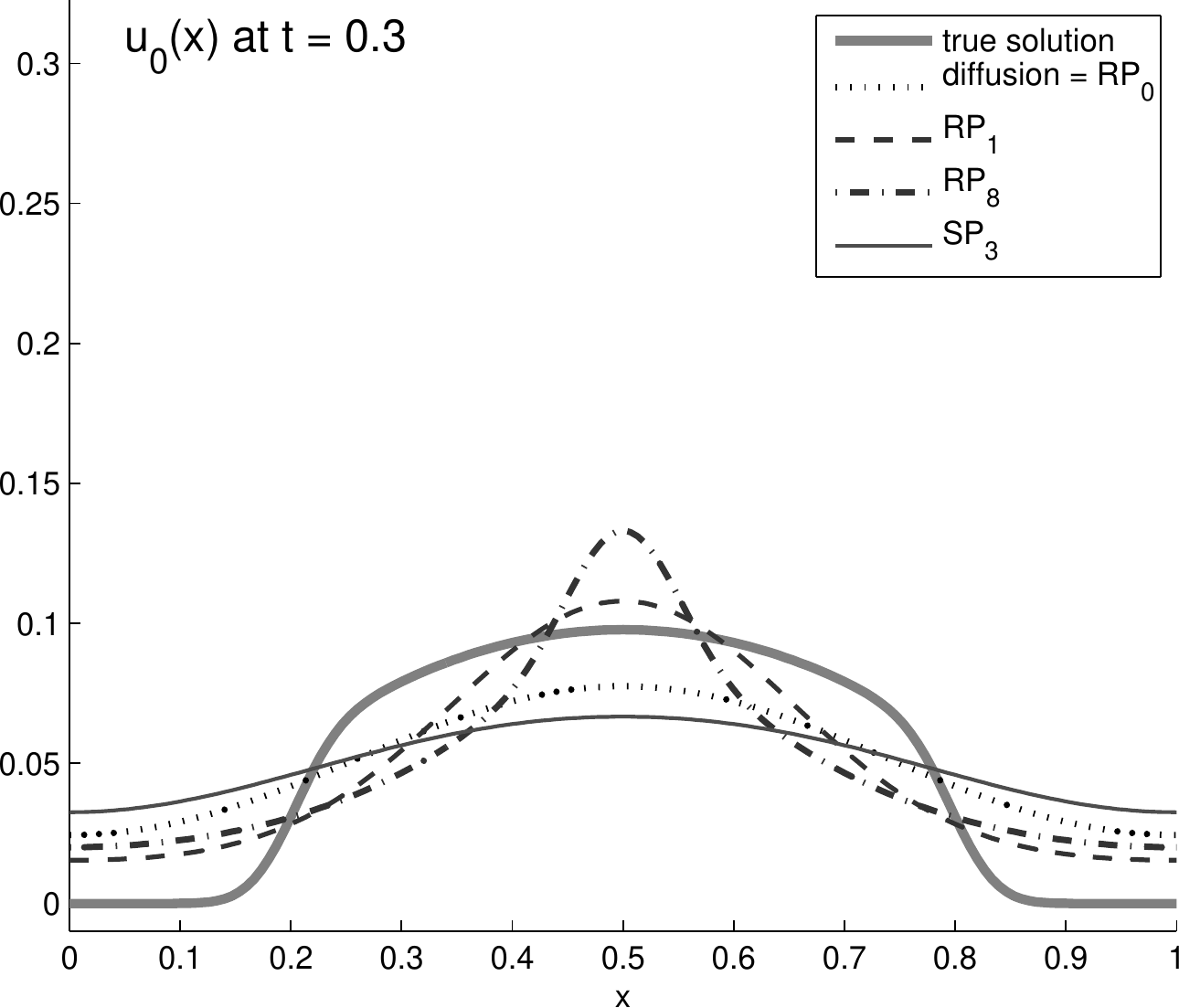}
\end{minipage}
\hfill
\begin{minipage}[t]{.49\textwidth}
\includegraphics[width=0.99\textwidth]{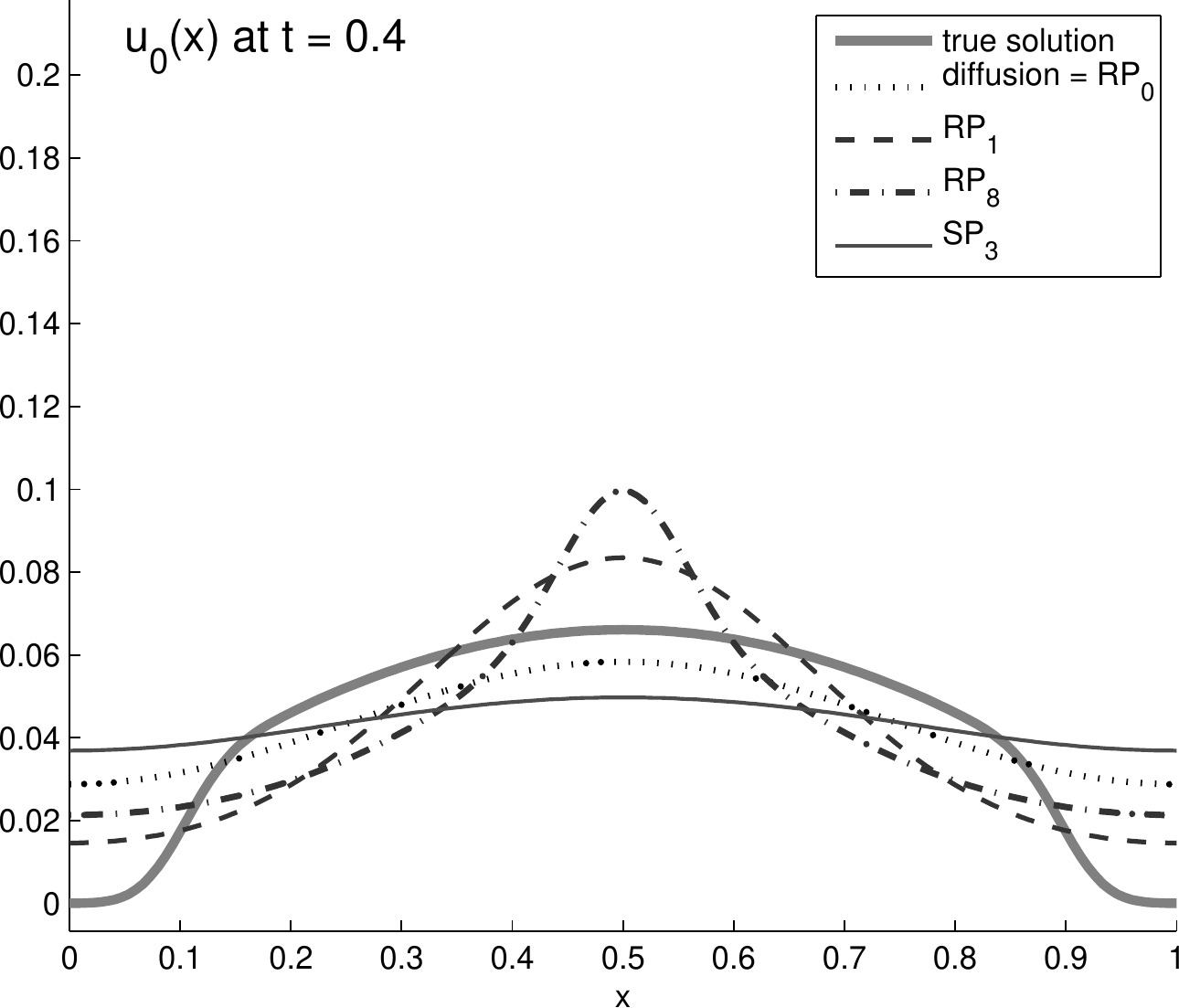}
\end{minipage}
\caption{Evolution of energy distribution for various parabolic approximations}
\label{fig:radiation_1d_parabolic}
\vspace{2em}
\centering
\begin{minipage}[t]{.49\textwidth}
\includegraphics[width=0.99\textwidth]{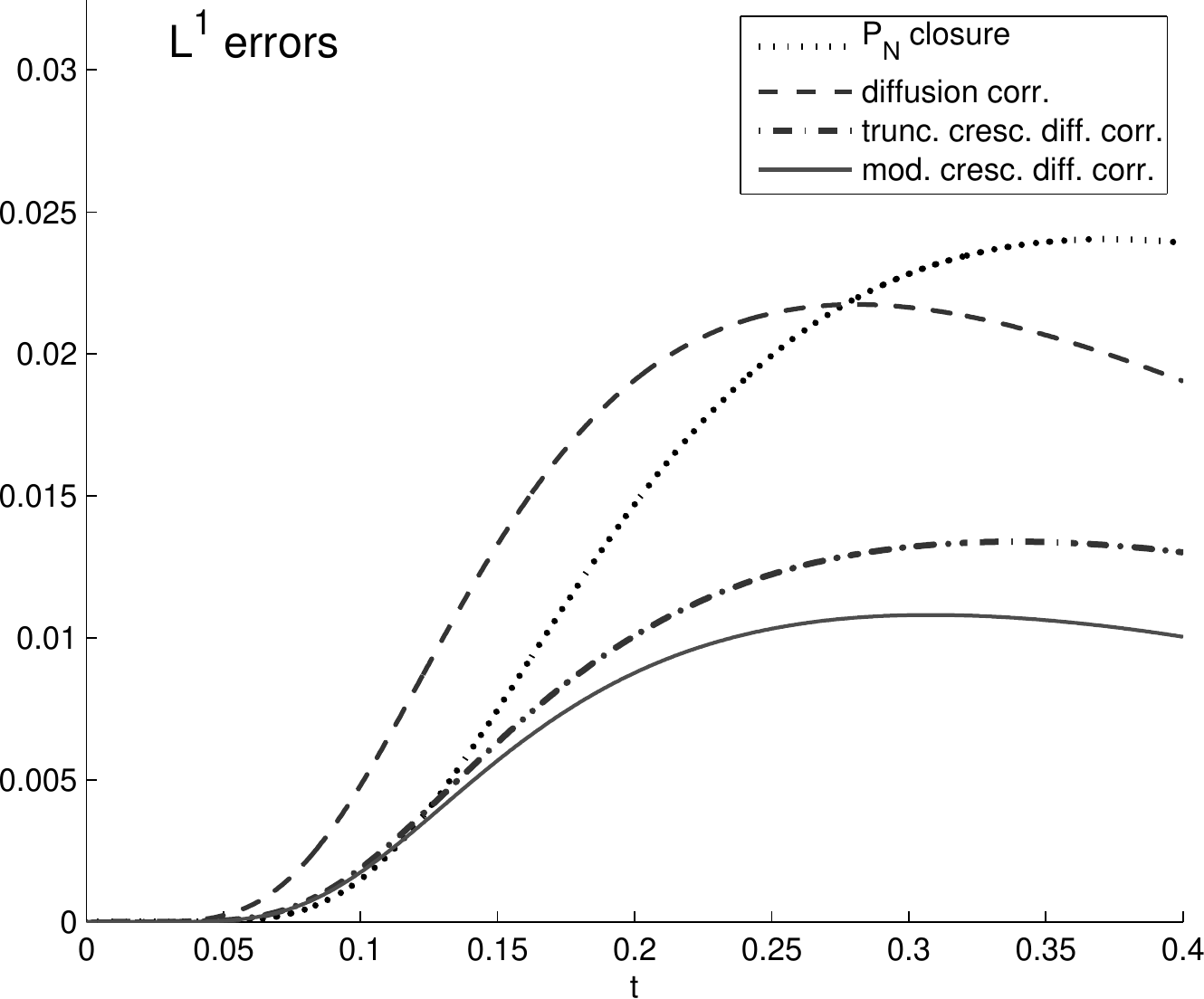}
\caption{Evolution of the $L^1$ error for moment closures for $N=3$}
\label{fig:radiation_1d_N3_error}
\end{minipage}
\hfill
\begin{minipage}[t]{.49\textwidth}
\includegraphics[width=0.99\textwidth]{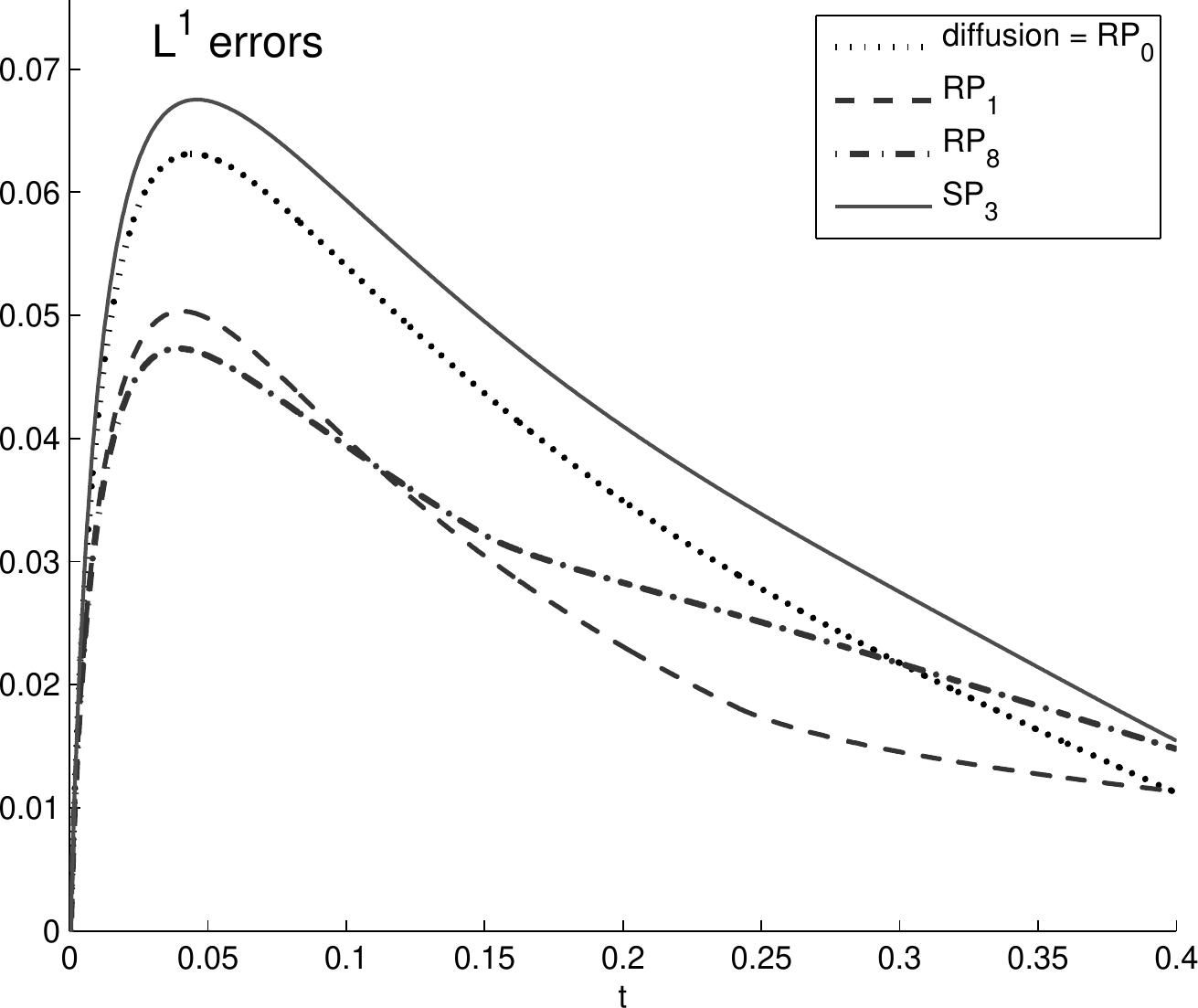}
\caption{Evolution of the $L^1$ error for parabolic approximations}
\label{fig:radiation_1d_parabolic_error}
\end{minipage}
\end{figure}

For $N=0$, the $P_N$ closure gives a sole exponential decay of the initial conditions,
which is a very bad approximation.
The diffusion closure is significantly better than $P_N$, but too diffusive.
Both versions of crescendo diffusion remedy this problem fairly well, and yield good
approximations, especially for short times. Note that here the truncated crescendo
diffusion reaches its full diffusivity at $t = \frac{1}{3}$.

For $N\ge 1$, the $P_N$ solution splits into $N$ peaks which move sideways with different
velocities. This is due to the fact that the $P_N$ model is a linear system of
hyperbolic equations with fixed wave speeds. With $N$ increasing, the height of the peaks
goes to zero, and the solutions approach the true solution.
The diffusive closures smear out the highest considered moment $u_N$, thus making it
more uniform and reducing its influence on the second-highest moment $u_{N-1}$.
Hence, qualitatively an $N$-th order diffusion correction solution lies ``between''
the $P_{N-1}$ and the $P_N$ solution. The results shown in
Fig.~\ref{fig:radiation_1d_N1} and Fig.~\ref{fig:radiation_1d_N3} visualize this effect.
In particular, one can observe that the classical diffusion correction is too close
to $P_{N-1}$, i.e.~the diffusion in the $N$-th moment is too large.
In contrast, the crescendo diffusion correction approximations ameliorate this effect.
The peak height is reduced, and the approximate solutions are closer to the truth.

The quality of the various approximations can be quantified by considering the
time evolution of the $L^1$-error of the approximate radiative intensity with respect
to the true solution
\begin{equation*}
e(t) = \int_0^1 u_0^\text{approx}(x,t)-u_0^\text{true}(x,t)\ud{x}\;.
\end{equation*}
The error of the various approximations is shown
in Fig.~\ref{fig:radiation_1d_N0_error} for $N=0$,
in Fig.~\ref{fig:radiation_1d_N1_error} for $N=1$, and
in Fig.~\ref{fig:radiation_1d_N3_error} for $N=3$.
One can observe that the errors become smaller as $N$ increases.
Within each figure, the $P_N$ closure yields the largest error at the final
time $t=0.4$. In contrast, the diffusion (correction) closures show smaller
errors at $t=0.4$. However, for short times, the error is increased compared to the
$P_N$ closure. This initial layer in time is particularly evident for $N=0$, shown
in Fig.~\ref{fig:radiation_1d_N0_error}.

Both versions of crescendo diffusion (correction) remove the initial layer of the
diffusion closure, and the following increase in error is significantly smaller than
with the $P_N$ closure. In terms of accuracy, no significant difference can be
observed between the two versions.

\subsection{Parabolic Approximations}
\label{subsec:numerical_results_parabolic}
Fig.~\ref{fig:radiation_1d_parabolic} shows a similar comparison for
parabolic type approximations, as derived in
Sect.~\ref{subsec:apply_op_radiation_reordered}.
The thick solid curve is the true solution. The dotted function is the classical
diffusion approximation, which equals the $RP_0$ system. The dashed curve shows the
$RP_1$ system \eqref{eq:RPN}, and the dash-dotted curve shows the $RP_8$ system.
For increasing order, the reordered $P_N$ equations converge to some limiting solution
that is not equal to the true solution. While this is easily explained by the fact that
the odd moments are not resolved, it implies that these systems are mainly useful for
small values of $N$. For comparison, the thin solid curve shows the
$SP_3$ system (with $\alpha=\tfrac{1}{3})$. One can observe that both the diffusion
approximations and the $SP_3$ equations are too diffusive. The $RP_N$ equations are far
from perfect, but they do yield better approximations.

The time evolution of the $L^1$ errors is shown
in Fig.~\ref{fig:radiation_1d_parabolic_error}.
For the considered examples, the errors obtained by the $RP_N$ equations are slightly
smaller than the errors obtained by the diffusion closure or by the $SP_3$.

\section{Conclusions and Outlook}
We have applied the method of optimal prediction to the infinite moment systems that
model radiative transfer. An integro-differential identity for the evolution of a finite
number of moments is obtained. The memory kernel involves an orthogonal dynamics
evolution operator. Simple approximations of this operator and of the integral have been
presented, and various approximate systems been derived. While traditionally closures
had been derived using physical arguments or by asymptotic analysis, the optimal
prediction formalism provides a very different strategy in which closures are derived
from a mathematical identity.

Using this methodology, existing closures can be re-derived, such as classical $P_N$
closures, diffusion and diffusion correction closures. In addition, new closures have
been obtained. Fundamentally new aspects are crescendo diffusion correction closures,
that modify classical diffusion correction closures by turning on the diffusion gradually
with time. In the context of optimal prediction, this explicit time dependence has a
natural interpretation as loss of information. While crescendo diffusion comes at no
additional cost, numerical tests indicate that the quality of the results is generally
improved.

In addition, the application of the formalism to a reordered version of the moment
system yields approximations of parabolic nature, which we denote
\emph{reordered $P_N$ equations}. Some versions of the existing simplified $P_N$
equations can be re-derived, and other systems can be obtained. A pleasing feature
of of these new $RP_N$ equations is that the diffusion matrices are always positive
definite.

Future research directions will focus on better approximations of the orthogonal
dynamics operator, as well as of the memory integral. Approximations of the orthogonal
dynamics can be based on the fact that higher order moments possess a uniform decay rate.
The memory integral can be approximated using the solution at former times, thus leading
to delay equations. While such systems pose numerical challenges, it may be very
rewarding to significantly improve approximations for very few considered moments.
Another limitation of the presented analysis lies in its linearity. Nonlinear
approximations, such as flux-limited diffusion \cite{Levermore1984} and minimum
entropy closures \cite{MullerRuggeri1993,AnilePennisiSammartino1991,DubrocaFeugeas1999}
would require nonlinear projections. While the optimal prediction formalism is not
restricted to linear projections, a precise formulation in the case of partial
differential equations is yet to be presented.

\bibliographystyle{amsplain}
\bibliography{references_complete}
\end{document}